\documentclass[10pt,aps,prd,superscriptaddress,twocolumn,nofootinbib]{revtex4-2}
\usepackage[paper=letterpaper,margin=1in]{geometry}
\usepackage[utf8]{inputenc}
\usepackage{dsfont}
\usepackage[dvipsnames]{xcolor}
\usepackage{amsfonts,amsmath,amssymb}
\allowdisplaybreaks[4]
\usepackage{euscript}
\usepackage{braket}
\usepackage{starfont}
\usepackage{color,soul}  
\usepackage{float}
\usepackage{tensor}
\usepackage{amsthm}
\usepackage{graphicx}
\usepackage{slashed}
\usepackage{leftidx}
\usepackage{subfigure}
\usepackage{bm}
\usepackage{bbm}
\usepackage{enumitem}
\definecolor{outerspace}{rgb}{0.25, 0.29, 0.3}
\definecolor{scarlet}{rgb}{1.0, 0.13, 0.0}
\usepackage[header,title,page,titletoc]{appendix}  
\definecolor{princetonorange}{rgb}{1.0, 0.56, 0.0}
\definecolor{WildStrawberry}{rgb}{1.0, 0.26, 0.64}
\definecolor{rossocorsa}{rgb}{0.83, 0.0, 0.0}
\definecolor{navyblue}{rgb}{0.0, 0.0, 0.5}
\usepackage{comment}
\usepackage{cancel}
\usepackage[normalem]{ulem}

\usepackage{mathtools}

\makeatletter
\newcommand{\dal}{\mathop{\mathpalette\dal@\relax}}
\newcommand{\dal@}[2]{%
  \begingroup
  \sbox\z@{$\m@th#1\square$}%
  \dimen0=\fontdimen8
    \ifx#1\displaystyle\textfont\else
    \ifx#1\textstyle\textfont\else
    \ifx#1\scriptstyle\scriptfont\else
    \scriptscriptfont\fi\fi\fi3
  \makebox[\wd\z@]{%
    \hbox to \ht\z@{%
      \vrule width \dimen0
      \kern-\dimen0
      \vbox to \ht\z@{
        \hrule height \dimen0 width \ht\z@
        \vss
        \hrule height 2\dimen0
      }%
      \kern-2.5\dimen0
      \vrule width 2.5\dimen0
    }%
  }%
  \endgroup
}

\makeatother

\usepackage{upgreek}

\newcommand{\dif}{\mathrm{d}} 
\newcommand{\talpha}{\widetilde{\alpha}} 
\newcommand{\M}{\mathsf{M}} 
\newcommand{\m}{\mathsf{m}} 
\newcommand{\p}{\mathsf{p}} 
\newcommand{\N}{\mathrm{N}} 
\newcommand{\GN}{G_\mathrm{N}} 
\newcommand{\rS}{r_\mathrm{S}} 
\newcommand{\rh}{r_\mathrm{h}} 
\newcommand{\tp}{t_\mathrm{p}} 
\newcommand{\rp}{r_\mathrm{p}} 
\newcommand{\Rp}{R_\mathrm{p}} 
\newcommand{\pc}{\mathsf{p}_\mathrm{c}} 
\newcommand{\Rmin}{R_\mathrm{min}} 
\newcommand{\nmax}{{n_\mathrm{max}}} 
\newcommand{\brho}{\bar{\rho}} 
\newcommand{\vrho}{\varrho} 
\newcommand{\bvrho}{\bar{\varrho}} 
\newcommand{\eff}{^\mathrm{eff}} 
\newcommand{\crit}{\mathrm{cr}} 
\newcommand{\ce}{\mathrm{c}}
\newcommand{\bo}{\mathrm{b}}

\newcommand{\be}{\begin{equation}}
\newcommand{\ee}{\end{equation}}

\usepackage{wasysym}
\usepackage{hyperref}
\hypersetup{
    colorlinks,
    citecolor=Peach,
    filecolor=Peach,
    linkcolor=MidnightBlue,
    urlcolor=MidnightBlue
}

\begin{document}

\title{Buchdahl limits in theories with regular black holes}

\author{Pablo Bueno}
\email{pablobueno@ub.edu}
\affiliation{Departament de F\'isica Qu\`antica i Astrof\'isica, Institut de Ci\`encies del Cosmos\\
 Universitat de Barcelona, Mart\'i i Franqu\`es 1, E-08028 Barcelona, Spain}

\author{Robie A. Hennigar}
\email{robie.a.hennigar@durham.ac.uk}
\affiliation{Centre for Particle Theory, Department of Mathematical Sciences,\\ Durham University, Durham DH1 3LE, UK}

\author{\'Angel J. Murcia}
\email{angelmurcia@icc.ub.edu}
\affiliation{Departament de F\'isica Qu\`antica i Astrof\'isica, Institut de Ci\`encies del Cosmos\\
 Universitat de Barcelona, Mart\'i i Franqu\`es 1, E-08028 Barcelona, Spain}

\author{Aitor Vicente-Cano}
\email{avicentecano@icc.ub.edu}
\affiliation{Departament de F\'isica Qu\`antica i Astrof\'isica, Institut de Ci\`encies del Cosmos\\
 Universitat de Barcelona, Mart\'i i Franqu\`es 1, E-08028 Barcelona, Spain}


\begin{abstract}
We study generalizations of Buchdahl's compactness limits for perfect-fluid star solutions of $D$-dimensional Einstein gravity coupled to higher-curvature corrections. We focus on Quasi-topological theories involving infinite towers of terms for which the unique vacuum spherically symmetric solutions correspond to regular black holes. We solve analytically the problem of constant-density stars and find that the space of solutions is bounded by: configurations with divergent central-pressure, corresponding to the most compact stars; 
configurations which possess zero central-pressure; and configurations for which the sizes of the stars coincide with the inner-horizon radii of the would-be regular black holes. In the more general case of perfect-fluid stars for which the mean density decreases with increasing radius, we show that, for each density profile, maximum compactness is reached when the metric becomes singular at the center. Under certain additional conditions, we find a novel Buchdahl limit for the maximum compactness of stars, attained by a specific constant-density profile. We show, in particular, that stars in these theories may be more compact than in Einstein gravity. While the vacuum solutions of these theories are such that all curvature invariants take mass-independent maximum finite values, we argue that there exist ordinary matter stars with finite central pressures for which such bounds can be violated---namely, arbitrarily high curvatures can be reached---unless additional constraints, such as the dominant energy condition, are imposed on the fluid. 
\end{abstract}
\onecolumngrid
\maketitle

\twocolumngrid


\section{Introduction}
In the context of general relativity (GR), Buchdahl’s theorem establishes a general bound on the maximum compactness attainable by perfect-fluid, isotropic stars with positive, outward-decreasing energy densities~\cite{PhysRev.116.1027}. According to this result, the areal radius $R$ of any such star of mass $M$ compatible with a finite central pressure must satisfy 
%
\begin{equation}
  R>R_{\rm Buch.} \qquad \text{where}\qquad  \frac{R_{\rm Buch.}}{2\GN M}=\frac{9}{8}\, ,
\end{equation}
which is slightly but significantly larger than the star's Schwarzschild radius. Remarkably, this bound coincides with the limit for constant-density stars, which  was actually found much earlier by Schwarzschild himself in~\cite{Schwarzschild:1916ae}---see also the related early works by Tolman~\cite{Tolman1934-TOLRTA,Tolman:1939jz}.

Numerous generalizations of this result have been obtained throughout the years. In particular, alternative maximum compactness limits follow from relaxing the isotropy and/or outward-decreasing density conditions~\cite{Bondi:1964zz,Florides,Guven:1999wm,Boehmer:2006ye,Ivanov:2002xf,Karageorgis:2007cy,Andreasson:2007ck,Arrechea:2024vxp}, from considering charged matter~\cite{deFelice:1995vkk,Mak:2001ie,Andreasson:2008xw,Shaymatov:2022hvs}, from taking into account the effects of rotation~\cite{Chakraborty:2022jgl} and from including semiclassical effects~\cite{Arrechea:2021vqj,Arrechea:2021xkp,Arrechea:2023oax,ElHanafy:2025raz}. Modified Buchdahl limits have also been identified for $D$-dimensional GR~\cite{PoncedeLeon:2000pj,Wright:2015dma} as well as for various modified gravity models~\cite{Olmo:2019flu,Kumar:2021vqa,Fernandez:2025cfp,Boehmer:2025uoq,Soranidis:2025arg}---in particular for Einstein-Gauss-Bonnet (EGB)\cite{Wright:2015yda,Bhar:2016fne,Malaver:2023ftk} and single Lovelock theories~\cite{Dadhich:2016fku,Dadhich:2015rea}.

In this paper we generalize Buchdahl's results to infinite families of higher-curvature modifications of GR in $D\geq 4$ whose most general spherically symmetric (SS) vacuum solutions are regular black holes. The theories under consideration belong to the so-called \emph{Quasi-topological} (QT) class~\cite{Oliva:2010eb,Quasi,Dehghani:2011vu,Ahmed:2017jod,Cisterna:2017umf,Bueno:2019ycr,Bueno:2022res,Moreno:2023rfl,Moreno:2023arp}, characterized by possessing second-order equations on spherically symmetric backgrounds.\footnote{In order for such theories to exist in $D=4$ one needs to resort to non-polynomial densities~\cite{Bueno:2025zaj}.} In these theories, as increasingly higher-curvature terms are included in the action, the divergences of the curvature invariants of the corresponding spherically symmetric black holes get milder. When arbitrarily high-curvature terms are considered, the singularity is fully resolved (as long as certain mild qualitative conditions on the gravitational couplings hold)~\cite{Bueno:2024dgm, Hennigar:2020kqt}. QT theories satisfy a Birkhoff theorem~\cite{Bueno:2025qjk}, so their most general SS vacuum solutions describe regular black hole generalizations of the Schwarzschild metric. These results provide a dynamical framework for the study of matter collapse and regular black hole formation, which has recently been exploited in~\cite{Bueno:2024zsx,Bueno:2024eig,Bueno:2025gjg,Bueno:2025zaj} in the case of thin shells and dust stars.  

A natural question concerns static equilibrium configurations within these models. In GR, curvature invariants reach arbitrarily high values at the star center as the central pressure limit is approached. On the other hand, QT theories with regular black holes satisfy a version of Markov's limiting curvature hypothesis~\cite{Frolov:1989pf,PismaZhETF.36.214}, namely, for vacuum solutions the magnitude of curvature invariants is bounded above by universal mass-independent quantities~\cite{Frolov:2024hhe,Bueno:2024zsx}. Hence, it is natural to inquire about the status of compactness limits in this context. Namely, one wonders whether an analogous mechanism prevents perfect-fluid stars from achieving arbitrarily high curvatures as they become increasingly compact.  As we show here, this is not the case, which suggests that less naive couplings between matter and gravity should be considered for these models in order for general curvature bounds to hold beyond the vacuum sector. Along the way, we unveil a rich structure of perfect-fluid star configurations which generalizes the GR results.

The paper is organized as follows. In Section~\ref{pfs} we explain our assumptions for the stellar matter considered, namely, a perfect fluid stress-tensor minimally coupled to our models. In Section~\ref{pfsE} we review the static and spherically symmetric (SSS) sector of $D$-dimensional GR, the problem of constant density stars, as well as the derivation of Buchdahl's inequalities for outward-decreasing density stars. We generalize these results for QT theories in Section~\ref{pfsQT}: we examine their SSS solutions in the presence of a perfect fluid, analyze constant density stars and obtain the corresponding Buchdahl limit for the maximum compactness of stars with outward-decreasing densities. 
We also explore under what conditions Markov's limiting curvature conjecture is satisfied by generic stars in QT theories. We present our conclusions and discuss future directions in Section~\ref{sec:disc}. In Appendix~\ref{app:photonring} we compute the size of QT regular black hole photon spheres and compare it to the stars compactness limits.

\section{Perfect fluid stars}\label{pfs}

In this section we review a few general aspects of perfect-fluid static and spherically symmetric $D$-dimensional stars. All the theories considered in this paper satisfy Birkhoff's theorem, which means that the exterior region will always be described by the corresponding  static (and generally unique) vacuum solution. On the other hand, the stress tensor will be non-vanishing within the matter-filled interior region which will be glued to the vacuum solution at the star surface---in the sense that the metric is continuous across the boundary, with no surface stress tensor.


In Schwarzschild coordinates, we can write the most general $D$-dimensional SSS metric as
%
\begin{equation}\label{Nf}
    \dif s^2 = - N(r)^2f(r) \dif t^2 + \frac{\dif r^2}{f(r)} + r^2\dif\Omega_{D-2}^2 \,,
\end{equation}
where $N(r)$ and $f(r)$ are two functions determined by the field equations, and $r^2\dif \Omega_{D-2}^2$ is the metric of a ($D-2$)-sphere of areal radius $r$, which henceforth will be referred to as the \textit{radius}.

We assume the matter to be described by a perfect fluid of mass-energy density $\rho(r)$, isotropic pressure $p(r)$ and equation of state $p(\rho)$.
The stress-energy tensor reads then
\begin{equation}\label{tab}
    T_{ab} = (\rho + p)u_a u_b + pg_{ab} \,,
\end{equation}
where $u^a$ is the fluid $D$-velocity. The staticity condition of the metric allows us to express the normalized velocity as $u^a=(N^2f)^{-1/2}\delta^a_t$. Thus, $T_{ab}$ is diagonal and its components read
\begin{equation} \label{eq_perfectfluid}
    T_{tt}\!=\!\rho N^2f \,,\  T_{rr}\!=\!\frac{p}{f} \,,\ T_{\theta_i\theta_i}\!=\!p\,r^2\!\prod_{j=1}^{i-1}\!\sin^2(\theta_j) \,,
\end{equation}
where there are $(D-3)$ polar angles $\theta_i\in[0,\pi]$, for $i=1,2,\dots,(D-3)$, and one azimuthal angle $\theta_{D-2}\in[0,2\pi)$.

The star is described by four functions of the radius $r$: the two metric functions, the density and the pressure. Thus, the structure of the star is determined by the two independent field equations, the local conservation law of the energy-momentum tensor $\nabla_{a}T^{ab}=0$, and the equation of state $p(\rho)$, along with the boundary conditions. The first two equations depend on the gravitational theory in which we are working, so we will consider them in the following sections. On the other hand, the expression for the conservation law can be obtained in a theory-independent  fashion. It reads
\begin{align} \label{eq_conservationlaw}
    \frac{\dif p}{\dif r} 
    = -\frac{1}{2}(\rho+p)\frac{\dif\log(N^2f)}{\dif r}\, .
\end{align}
Finally, the boundary conditions must ensure the continuity of the pressure and the metric at the surface of the star, which we define as $r=R$. 

Let us now introduce a few functions that will be useful in order to describe a star of Arnowitt-Deser-Misner (ADM) mass $M$ and radius $R$. First of all, let us define the mass inside a sphere of radius $r$ by
\begin{equation}
    m(r) \equiv  \Omega_{D-2} \int_0^r\dif x \, x^{D-2} \rho(x) \,,
\end{equation}
where
\begin{equation}
    \Omega_{D-2}\equiv \frac{2\pi^{(D-1)/2}}{\Gamma[\frac{D-1}{2}]} \,,
\end{equation}
is the volume of a ($D-2$)-sphere. 
The total mass is the value of the previous mass function at the star's radius, $m(R)=M$. As it turns out, it is often more convenient to work with the mean density within a sphere of radius $r$.
Let us thus define
\begin{equation}
    \brho(r) \equiv \frac{(D-1)m(r)}{\Omega_{D-2}r^{D-1}} \,.
\end{equation}
Similarly, in order to minimize the repeated use of specific constants, let us define reduced versions of the relevant magnitudes, as follows,
\begin{align}\label{defis}
   \{ \m,\M \} & \equiv \frac{8\pi\GN }{(D-2)\Omega_{D-2}} \{m,M \} \,,
   \\
   \{ \vrho, \bvrho,\p \} & \equiv  \frac{8\pi\GN }{(D-2)(D-1)} \{\rho,\brho,p \} \,,
\end{align}
where $\GN$ is Newton's constant. Observe that one can express the density $\vrho$ in terms of the mean density $\bvrho$ by differentiating
\begin{equation}\label{rhom}
    \vrho = \bvrho + \frac{r}{D-1}\bvrho_{,r} \,,
\end{equation}
where the subscript following a comma denotes differentiation with respect to that variable (in this case, $r$). Additionally, the mass defines a length parameter corresponding to the Schwarzschild radius of the star, given by $\rS \equiv  \left(2\M\right)^{1/(D-3)}$.


\section{Stars in Einstein gravity}\label{pfsE}
Let us start with the case of GR. The $D$-dimensional Einstein-Hilbert action minimally coupled to matter reads
\begin{equation}
    I=\int\dif^D x \sqrt{|g|} \left[\frac{R}{16\pi \GN}+\mathcal{L}_{\rm matter} \right]\, .
\end{equation}
The perfect-fluid stress tensor~\eqref{tab} can be obtained from an explicit matter Lagrangian like the one introduced above in the standard way, namely, 
\begin{equation}
    T_{ab} = -\frac{2}{\sqrt{|g|}}\frac{\delta I_{\rm matter}}{\delta g^{ab}} \,,
\end{equation}
although constructing $\mathcal{L}_{\rm matter}$ turns out to be trickier than naively expected~\cite{Brown:1992kc}. Varying the action with respect to the metric leads to Einstein equations, 
\begin{equation}
    R_{ab}-\frac{1}{2}g_{ab}R=8\pi \GN T_{ab} \,.
\end{equation}


\subsection{Equations of motion}
As it is well-known, Birkhoff's theorem holds in GR: the Schwarzschild-Tangherlini metric is the unique vacuum solution with spherical symmetry~\cite{Tangherlini:1963bw}. For our purposes, it suffices to consider a spherically symmetric ansatz which is static from the outset, namely,~\eqref{Nf}.  In the presence of matter, the non-vanishing components of the Einstein equations read
%
\begin{align} \label{eq_EinsteinEqs}
   &  \frac{\dif }{\dif r} \left[r^{D-3}(1-f)\right] = \frac{16\pi\GN}{(D-2)}\frac{r^{D-2}}{N^2f}T_{tt} \,, \\
   &   \frac{N_{,r}}{rN} = \frac{8\pi\GN}{(D-2)}\left[\frac{1}{N^2f^2}T_{tt} + T_{rr}\right] \,,
\end{align}
plus the angular components, which provide no additional constraints by virtue of the Bianchi identity, as long as the stress tensor is covariantly conserved. 

\subsubsection{Exterior solution}
In the absence of matter, the above equations can be straightforwardly solved. The result is
%
\begin{equation}
    f(r) = 1 - \frac{2\M}{r^{D-3}} \,, \quad N(r) = C \,,
\end{equation}
where $\M$ and $C$ are two integration constants. Naturally, $\M$ is proportional to the ADM mass of the spacetime through~\eqref{defis}, whereas we may set $C$ to $1$ after a trivial time rescaling, if desired. The result is the $D$-dimensional  Schwarzschild-Tangherlini solution. This describes the exterior of the star regardless of its matter content (and dynamics).



\subsubsection{Interior solution}
Let us consider now the interior of the star. 
By plugging~\eqref{eq_perfectfluid} into the Einstein equations~\eqref{eq_EinsteinEqs}, one finds the following equations for $f$ and $N$, 
\begin{align}
    f(r) & = 1-\frac{2\m(r)}{r^{D-3}} = 1 - 2\bvrho(r) r^2 \,, \\
    \frac{\dif \log N(r)}{\dif r} & = \frac{(D-1)r}{f(r)} [\vrho(r)+\p(r)] \,,
    \label{eq_logN}
\end{align}
where we momentarily made explicit all the dependencies on $r$. 

By combining the above equations with the conservation law of the stress-energy tensor~\eqref{eq_conservationlaw}, it is possible to derive two decoupled ODEs: one for the pressure and one for a combination of the metric functions. Inserting~\eqref{eq_logN} in~\eqref{eq_conservationlaw}, we obtain the $D$-dimensional version of the Tolman-Oppenheimer-Volkoff (TOV) equation for the hydrostatic equilibrium~\cite{PhysRev.55.374},
\begin{equation} \label{eq:TOV_GR}
    \p_{,r} = -\left(\vrho + \p\right)r\,\frac{(D-3)\bvrho + (D-1)\p}{f(r)} \,.
\end{equation}

At this stage, it will be convenient to introduce a new set of variables. First, let us define $x\equiv r^2$ and the lapse function $\zeta\equiv N f^{1/2}$.
Then, introducing  $\xi$ from $\dif \xi \equiv  \dif x/\sqrt{f}$, the second decoupled equation reads
\begin{equation}\label{eq_zeta_ode}
    \zeta_{,\xi\xi} = g(\xi) \zeta \,,
\end{equation}
where we defined
\begin{equation}\label{gxi_eq_GR}
    g(\xi)=\frac{(D-3)}{2}\,\bvrho_{,x} \,.
\end{equation}

The continuity conditions impose: $\m(R)=\M$, $N(R) = 1$ and $\p(R)=0$. Also, regularity at the center imposes $f(0)=1$.

\subsection{Constant-density stars}
The simplest stellar model corresponds to an incompressible equation of state---that is, to stars with constant density, for which
\begin{equation}
   \vrho=\bvrho=\frac{\M}{R^{D-1}}\equiv \vrho_0 \,,
\end{equation}
where $\vrho_0$ is constant.\footnote{This is a particular configuration within the family of Tolman type III solutions.}
In that case, the  metric function $f(r)$ takes a de Sitter-like form in the interior,\footnote{Since $f(r)^{-1}$ is the $rr$ component of the metric, one can obtain the proper length along the radial direction in star's interior from the above expression analytically. The result is $\rp = a \arcsin(r/a) \, ,$ where $a\equiv \sqrt{R^{D-1}/(2\M)}=R/\sqrt{2\vrho_0}$. Thus, an injective mapping between both variables holds. However, the proper radius $\Rp$ and the areal radius of the star $R$ are not related by an injective function.}
%
\begin{align}
    f(r) = 1 - \frac{2\M r^2}{R^{D-1}} = 1-2\vrho_0 r^2 \,.
\end{align}
Naturally, at the star surface this reduces to
\begin{equation}
     f(R) = 1 - \frac{2\M}{R^{D-3}} \,,
     \label{boundfgr}
\end{equation}
which matches the corresponding exterior-solution metric function.

The knowledge of the density profile and $f(r)$ allows to solve the TOV equation. The resulting pressure profile reads
\begin{equation} \label{pressure_GR}
   \frac{\p(r)}{\vrho_0} = \frac{1-\left(\dfrac{f(R)}{f(r)}\right)^{1/2}}{\dfrac{(D-1)}{(D-3)}\left(\dfrac{f(R)}{f(r)}\right)^{1/2}-1} \,.
\end{equation}
%
%
%
The central pressure, defined as $\pc=\p(0)$, follows straightforwardly and reads
\begin{equation}\label{centralp}
   \frac{\pc}{\vrho_0} = \frac{1-f(R)^{1/2}}{\dfrac{(D-1)}{(D-3)}f(R)^{1/2}-1} \,.
\end{equation}
%
%
%
Finally, the remaining metric function of the interior solution, $N(r)$, can be obtained by integrating~\eqref{eq_zeta_ode} together with Cauchy boundary conditions at $r=R$. The result reads\footnote{This results in the following relation between the coordinate time $t$ and the proper time $\tp$
\begin{equation}
    \frac{\dif\tp}{\dif t} =
    \begin{cases} 
     \frac{D-1}{2}f(R)^{1/2} - \frac{D-3}{2}f(r)^{1/2} & 0<r<R \\
    f(r)^{1/2}& r\geq R \,.
    \end{cases}
\end{equation}}
%
\begin{equation}\label{Nc}
    N(r) = \dfrac{(D-1)}{2} \left(\dfrac{f(R)}{f(r)}\right)^{1/2}-\dfrac{(D-3)}{2} \,,
\end{equation}
which satisfies $N(R)=1$ at the star surface. 

Additionally, observe that the pressure expression in~\eqref{pressure_GR} can be written in terms of the metric function $N$, using the result~\eqref{Nc} through
\begin{equation} \label{eq_GR_pN}
    \frac{\p(r)}{\vrho_0} = \frac{(D-3)}{(D-1)} \frac{1-N(r)}{N(r)} \,.
\end{equation}
This relation will help shed some light on the connection between the pressure and the metric in the context of maximum compactness bounds.

\begin{figure}[t!]
    \centering
    \includegraphics[width=0.48\textwidth]{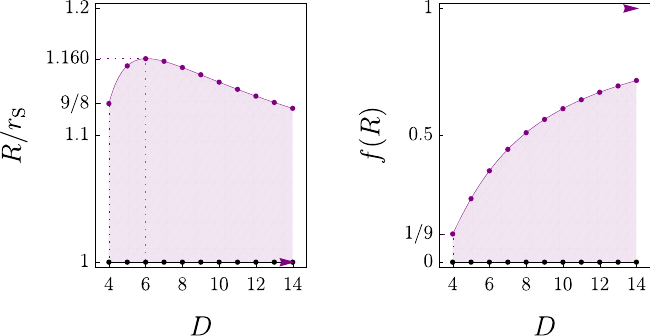}
    \caption{ {Maximum compactness limit for constant-density stars in Einstein gravity as a function of the spacetime dimension. Purple dots indicate the value of the minimal radius (infinite-central-pressure), black dots for black holes, arrowheads for the limit as the dimension approaches to infinity, and we shade the regions accordingly: white for regular stars, and purple for stars that are physically unacceptable because the pressure $\p(r)$ diverges at some interior point $r$. (Left) Minimal radius divided by the Schwarzschild radius. We observe that in $D=6$, this quantity reaches its maximum value, and the ratio saturates as $D\to\infty$. (Right) Metric function at the limiting star's surface $f(\Rmin)$. This value increases monotonically with $D$, so it reaches its minimum at $D=4$, meaning that the gravitational effects are strongest in four dimensions. Likewise, for large $D$, the metric function approaches 1.}}
    \label{fig_GR_Buchdahl}
\end{figure}
\begin{figure}[h!]
    \centering
    \includegraphics[width=0.48\textwidth]{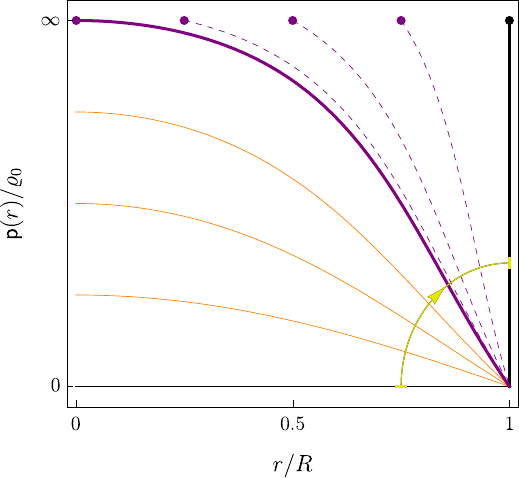}
    \caption{
    {We plot a family of pressure profiles for different star parameters in $D=6$ Einstein gravity (the plot is qualitatively analogous $\forall\,D\neq 6$). The yellow arrow indicates the direction in which the star radius $R$ decreases, starting from a pressureless star with a very large radius (X-axis), all the way to an unphysical ultracompact star with a Schwarzschild radius (black vertical line). Orange lines correspond to physical stars and purple dashed lines to unphysical ones. The limiting case of a star saturating the compactness  limit is represented by a solid purple line. This has a point-like singularity at $r=0$ (indicated with a dot).The purple dashed lines represent would-be finite-volume singularities. 
    }}
    \label{fig_profilesGR}
\end{figure}

\subsubsection{Maximum compactness limit}
The formula for the central pressure allows us to compute the maximum compactness limit for constant-density stars with a non-divergent-central-pressure, or, in other words, with a non-degenerate metric in $D$ dimensions. 

This limiting singular case, for which the curvature invariants diverge, takes place when the radius of the star reaches the minimum value
%
%
%
\begin{equation}\label{buchiE}
    \frac{\Rmin}{\rS} = \left[\frac{(D-1)^2}{4(D-2)}\right]^{1/(D-3)} \,,
\end{equation}
which we expressed as a fraction of the Schwarzschild radius of the star $\rS$. 
The classical result of $9/8$ is obtained by setting $D=4$~\cite{Schwarzschild:1916ae}. Furthermore, as shown in Fig.~\ref{fig_GR_Buchdahl},  the limit reaches the lowest-compactness value for $D=6$, and it saturates as $D$ approaches infinity. In that limit, $\Rmin$ coincides with the Schwarzschild radius, which means that arbitrarily compact stars  are possible for $D\rightarrow +\infty$, provided their radius is greater than $\rS$. Alternatively, since $R$ and $f(R)$ are related by an injective relation inherited by the Schwarzschild-Tangherlini solution, the minimum value that the blackening factor takes at the star's surface is given by~\cite{PoncedeLeon:2000pj} 
\begin{equation}\label{redsh}
  f(\Rmin) = 1-\left(\frac{\rS}{\Rmin}\right)^{D-3} = \left(\frac{D-3}{D-1}\right)^2 \,.
\end{equation}
As shown in Fig.~\ref{fig_GR_Buchdahl}, its maximum compactness occurs in 4 dimensions and increases with the number of dimensions, saturating at 1 as $D\to+\infty$. 

For later comparison, in Fig.~\ref{fig_profilesGR} we have plotted the pressure profiles of various Einstein gravity constant-density stars in $D=6$. In order to display the exact point at which the pressure becomes infinite for stars exceeding the maximum compactness bound, we compactify the Y-axis using an $\arctan$ function. Hence, when the pressure reaches the top of the diagram, such a star would involve a singularity at the radial coordinate point marked with a dot.

\subsection{Buchdahl's inequalities}
We have just seen that constant-density stars are such that there exists a maximum possible degree of compactness compatible with a non-singular star configuration. Following Buchdahl's original argument~\cite{PhysRev.116.1027}, let us now analyze the same problem in the case of a general equation of state under the (necessary) assumption that the mean mass density of the star does not increase as we move outward.\footnote{It is worth noting that the original (sufficient) assumption based on ``physically reasonable stars'' is more restrictive, as it refers to the local density $\vrho$ rather the mean density $\bvrho$. However, the derivation of the compactness limits follows the same procedure.}


In order to tackle this problem, we make use of the second-order ODE~\eqref{eq_zeta_ode} for the lapse function $\zeta$. 
%
%
%
%
Since $g(\xi)$ is negative by hypothesis, we have
\begin{equation}\label{desig}
    (\zeta_{,\xi})_{\rm c} \geq \zeta_{,\xi} \geq (\zeta_{,\xi})_{\rm b} \,,
\end{equation}
where the subscripts ``c'' and ``b'' represent evaluation at the center and at the boundary of the star, respectively.

At the star boundary $r=R$, the metric functions reduce to their vacuum form, so we can evaluate explicitly this quantity,
\begin{equation}
    (\zeta_{,\xi})_{\rm b} = \left.\frac{\dif \sqrt{f}}{\dif x/\sqrt{f}}\right|_{r=R}  = \frac{(D-3)}{2}\bvrho_{\rm b} \,.
\end{equation}
And from the second inequality in~\eqref{desig}, one finds
%
\begin{equation}\label{ini1}
    \zeta_{,\xi}\dif \xi  \geq \frac{(D-3)}{2}\bvrho_{\rm b} \dif \xi \,.
\end{equation}
Trivially integrating the left-hand side (LHS), we have
\begin{equation}\label{io}
    \int_{\rm c}^{\rm b} \zeta_{,\xi} \dif\xi  = \zeta_{\rm b}-\zeta_{\rm c}=\sqrt{f(R)}-N(0) \,,
\end{equation}
where we used that $f(0)=1$ and $N(R)=1$.
On the other hand, on the right-hand side (RHS) we have
\begin{align}
  \int_{\rm c}^{\rm b}  \bvrho_{\rm b}  \dif \xi = \int_{\rm c}^{\rm b} \frac{2\bvrho_{\rm b} r\dif r}{\sqrt{f}}=\int_{\rm c}^{\rm b}  \frac{2 \bvrho_{\rm b}r\dif r}{\sqrt{1-2 \bvrho(r)r^2}} \, , 
\end{align}
where we omitted the $(D-3)/2$ factor. Now, taking into account that $\bvrho_{,r}\leq0$ by hypothesis, we have
\begin{equation}
    \frac{1}{\sqrt{1-2 \bvrho(r) r^2}} \geq \frac{1}{\sqrt{1-2 \bvrho_{\rm b}r^2}}
\end{equation}
$\forall r \in [0,R)$. Hence,
\begin{align}\label{op}
  \int_{\rm c}^{\rm b}  \bvrho_{\rm b}  \dif \xi \geq  \int_{\rm c}^{\rm b}  \frac{2 \bvrho_{\rm b}r\dif r}{\sqrt{1-2 \bvrho_{\rm b}r^2}}=1-\sqrt{f(R)} \,.
\end{align}
Combining~\eqref{ini1} with~\eqref{io} and~\eqref{op}, we finally have
%
%
\begin{equation}
   \sqrt{f(R)}-N(0) \geq \frac{(D-3)}{2} \left(1-\sqrt{f(R)}\right) \,.
\end{equation}
Rearranging this expression, we are left with
\begin{equation}\label{fN0}
   f(R) \geq \left(\frac{D-3+2N(0)}{D-1}\right)^2 \,, 
\end{equation}
%
Since well-behaved stars have a regular center $N(0)>0$, the limiting configuration for any density profile occurs at some $\Rmin$, when $N(0)=0$, that is, when the metric would describe a singularity. Consequently, one finds
\begin{equation}\label{fN01}
   f(\Rmin) \geq \left(\frac{D-3}{D-1}\right)^2 \,,
\end{equation}
where the equality in~\eqref{fN0} and~\eqref{fN01} holds only for constant-density stars. Therefore, any star with radius $R$ must obey
\begin{equation}
    f(R) > \left. f(R)\right|_\mathrm{min} \geq \left(\frac{D-3}{D-1}\right)^2 \,,
\end{equation}
or equivalently, as $\left. f(R)\right|_\mathrm{min} = \left. f(R_\mathrm{min}\right)$,
\begin{equation}
    \frac{R}{\rS} > \frac{R_{\rm min}}{\rS} \geq \left[\frac{(D-1)^2}{4(D-2)}\right]^{1/(D-3)} \equiv \frac{R_{\rm Buch.}}{\rS} \,.
\end{equation}
%
Observe that only configurations with constant density can attain the RHS bound. Therefore, we define a Buchdahl star as the limiting case of maximal compactness, corresponding to a star with constant density and radius $R_{\rm Buch.}$.

%
%
Hence, given the set of all possible stars with monotonically decreasing mean densities compatible with Einstein equations, the value of $f(R)$ is bounded below by 
\eqref{fN0}, which depends on the value of $N(r)$ at the center. The smallest possible value is reached within the divergent-central-pressure  stars subclass and, within those, for constant-density stars. 

\section{Stars in higher-curvature gravities}\label{pfsQT}
Let us now study the effect of introducing higher-curvature corrections to the Einstein-Hilbert action.

\subsection{Quasi-topological gravities}
Consider a general action built from contractions of the Riemann tensor and the metric,
\begin{equation}
    I=\int \dif^D x \sqrt{|g|} \left[\frac{\mathcal{L}\left (R_{abcd},g^{ef} \right)}{16\pi \GN}+\mathcal{L}_{\rm matter} \right] \,.
\end{equation}
The equations of motion read now~\cite{Padmanabhan:2011ex}
\begin{align}\label{eq:Eab}
    \!P_a{}^{cde}R_{bcde}\!-\!\frac{g_{ab}\mathcal{L}}{2}\!-\!2\nabla^c\nabla^d P_{acdb}\!=\!8\pi \GN T_{ab} \,,  
\end{align}
where we defined
\begin{equation}
    P^{abcd}\equiv \left[ \frac{\partial \mathcal{L}}{\partial R_{abcd}}\right]\, ,
\end{equation}
an object which has the same symmetries as the Riemann tensor by construction.
In general, the above equations are fourth-order in derivatives. However, demanding the third term in the LHS to be absent reduces the order to two and defines the class of theories known as ``Lovelock gravities''~\cite{Lovelock1,Lovelock2},
\begin{equation}
    \nabla^d P_{acdb}=0\quad \Leftrightarrow \quad \text{Lovelock gravity}\, .
\end{equation}
For a given spacetime dimension, Lovelock densities with non-trivial equations of motion exist at every curvature order $n$ up to  $\lfloor\frac{D-1}{2}\rfloor$. In even dimensions, the special case $n=D/2$ corresponds to topological densities which have trivial equations of motion but which do not vanish on-shell in general---their integral on a compact manifold is proportional to the Euler characteristic of the manifold. On the other hand, all Lovelock densities with $n> D/2$ are identically zero.

A broader class of theories which does contain non-vanishing representatives at every curvature order in general dimensions $D\geq 5$ is provided by Quasi-topological (QT) gravities\footnote{The single exception takes place at order $n=D/2$ in even dimensions, for which all QT invariants are trivial.}~\cite{Oliva:2010eb,Quasi,Dehghani:2011vu,Ahmed:2017jod,Cisterna:2017umf,Bueno:2019ycr, Bueno:2022res, Moreno:2023rfl,Moreno:2023arp}. There exist several slightly inequivalent definitions of QT gravities in the literature. As recently argued in~\cite{Bueno:2025qjk}, it appears that the most natural notion for a QT is the following one:
\begin{equation}
  \left.  \nabla^d P_{acdb}\right|_{N,f}=0\quad \Leftrightarrow \quad \text{QT gravity}\, ,
\end{equation}
namely, if the above tensorial structure vanishes identically when evaluated on the static and spherically symmetric ansatz~\eqref{Nf}. With the exception of the $n=1$ and $n=2$ cases, for which the only QT densities are the Einstein-Hilbert and Gauss-Bonnet terms, given a curvature order $n$ there exist several QT Lagrangians which yield the same equations for SSS metrics. 

Remarkably, non-trivial QT theories also exist at every curvature order (except for $n=D/2$) and for general $D \geq 5$. Except for a measure-zero subset of theories, QT gravities satisfy a Birkhoff theorem and their most general SS vacuum solutions are static and given by the corresponding generalization of the Schwarzschild-Tangherlini spacetime \cite{Bueno:2025qjk}. 

%
%
Their Lagrangian can be written as 
\begin{equation}\label{genAc}
   \mathcal{L}\left (R_{abcd},g^{ef} \right) = R+\!\sum_{n=2}^{n_{\rm max}}\alpha_n \mathcal{Z}_n \, ,
\end{equation}
where $\alpha_n$ are arbitrary couplings with dimensions of length$^{2(n-1)}$. Hence, they introduce at least one new length scale in the theory.
From now on, we will restrict our analysis to theories for which $\alpha_n(D-2n)\geq0$ $\forall n$. This choice simplifies certain calculations and it is also convenient in order for the theories to allow for simple regular generalizations of the Schwarzschild-Tangherlini black hole when $n_{\rm max}\rightarrow \infty$~\cite{Bueno:2024dgm}.\footnote{See~\cite{Konoplya:2024kih,Konoplya:2024hfg,DiFilippo:2024mwm,Ma:2024olw,Cisterna:2024ksz,Ditta:2024iky,Wang:2024zlq,Hennigar:2020kqt,Fernandes:2025eoc,Fernandes:2025fnz,Cisterna:2025vxk,Ling:2025ncw,Eichhorn:2025pgy,Fernandes:2025mic,Aguayo:2025xfi,Boyanov:2025pes,Konoplya:2025uta,Hao:2025utc,Frolov:2025ddw} for related works motivated by this construction.} The QT densities $\mathcal{Z}_n$ 
can be obtained at arbitrary order starting from the first five, $\mathcal{Z}_i$, $\{i=1,\dots, 5\}$, from the recursive formula~\cite{Bueno:2019ycr}
\begin{equation}
\begin{aligned}\label{recu}
\mathcal{Z}_{n+5}=
&\frac{3(n+3)\mathcal{Z}_{1}\mathcal{Z}_{n+4}}{D(D-1)(n+1)}-\frac{3(n+4)\mathcal{Z}_{2}\mathcal{Z}_{n+3}}{D(D-1)n}\\ &+\frac{(n+3)(n+4)\mathcal{Z}_{3}\mathcal{Z}_{n+2}}{D(D-1)n(n+1)}\, .
\end{aligned}
\end{equation}
The explicit form of the first five densities is not particularly illuminating and can be found, for instance in Eq.~7 of~\cite{Bueno:2024zsx}.

More recently, it has been shown that QT theories exist also in $D=4$ provided one relaxes the assumption that the densities must involve \emph{ polynomials} of the Riemann tensor~\cite{Bueno:2025zaj}---see also~\cite{Chinaglia:2017wim, Colleaux:2017ibe,Colleaux:2019ckh} for related previous works. Remarkably, a recursive formula identical to~\eqref{recu} can be used to obtain arbitrarily higher-order densities. The first five terms, which in this case involve quotients of polynomial densities, can be found in Eqs.\,1--5 of~\cite{Bueno:2025zaj}.


\subsection{Equations of motion} \label{sec:4.B}
As far as SS metrics are concerned, the Lovelock and Quasi-topological equations are identical, the only difference being the order at which the series~\eqref{genAc} is truncated---\emph{i.e.,} $\lfloor\frac{D-1}{2}\rfloor$ for Lovelock and arbitrarily large for QT. The non-vanishing components of the equations of motion on a general SSS ansatz read now \cite{Bueno:2024zsx}:
\begin{align} \label{eq_QTequations0}
    \frac{\dif}{\dif r}\left[r^{D-1}h(\psi)\right] = \frac{16\pi\GN}{(D-2)} \frac{r^{D-2}}{N^2f}T_{tt} \,, \\ \label{eq_QTequations1}
    \frac{N_{,r}}{rN}\!=\!\frac{8\pi\GN}{(D\!-\!2)h'(\psi)}\left(\!\frac{1}{N^2f^2}T_{tt} \!+\! T_{rr}\!\right) \,,
\end{align}
where $h'(\psi)\equiv\frac{\dif h(\psi)}{\dif\psi}$. These are remarkably similar to the Einstein gravity ones. All the information about the theory under consideration is encapsulated in the ``characteristic polynomial'', $h(\psi)$, which is defined as
\begin{equation}\label{char_poly}
h(\psi) \equiv \psi + \sum_{n=2}^\nmax \widetilde{\alpha}_n \psi^n\, ,  
\end{equation}
where in~\eqref{eq_QTequations0} we defined
\begin{equation} \label{eq_f_psi}
    \psi \equiv \frac{1-f(r)}{r^2}\, ,\qquad  \widetilde{\alpha}_n \equiv \alpha_n\frac{(D-2n)}{(D-2)}\, .
\end{equation}
%
%
%
For convenience, from now on we will refer to the $ \widetilde{\alpha}_n $ when making comments about the gravitational coefficients.

\subsubsection{Exterior solution}
In the absence of matter, the equations reduce to
\begin{equation}\label{high_vacuum}
    h(\psi)= \frac{2\M}{r^{D-1}} \,, \quad  \ N(r) = C \,,
\end{equation}
where $\M$ and $C$ are two integration constants. Just like for Einstein gravity, $\M$ is proportional to the ADM mass, and we will refer to it as the mass, while $C$ may be set to $1$ after a time rescaling.
The metric of the exterior of the star is therefore implicitly determined by $h(\psi)$ and the constant $\M$. Observe that the condition $\talpha_n\geq0$ implies that every derivative of $h(\psi)$ with respect to $\psi$ is a monotonically increasing function. Also, we maintain the notation $r_{\rm S}\equiv (2\M)^{1/(D-3)}$, although $r_{\rm S}$ will no longer represent the horizon radius of the putative black hole of mass $\M$ for the corresponding theory. 
%
%

To determine the number of horizons that these solutions can possess, we analyze the metric function which, using~\eqref{eq_f_psi} and~\eqref{high_vacuum}, can be written as
\begin{equation} \label{eq:new_metric}
    f(r) = 1 - \left(\frac{\rS}{r}\right)^{D-3}\frac{\psi}{h(\psi)} \,.
\end{equation}
Horizons correspond to the zeros of $f(r)$. Now, let us compute $f_{,r}$ and study whether there exists $r=r_{0}$ for which $f_{,r}(r_0)=0$:
\begin{equation}
\label{psihp}
f_{,r}(r_0)= 0 \quad   \Longrightarrow \quad  \left.\frac{\psi h'(\psi)}{h(\psi)}\right|_{r=r_0} \! = \frac{D-1}{2} \,.
\end{equation} 
Since $h(\psi)$ is monotonically increasing and $\talpha_n \geq 0$, \eqref{psihp} will have a unique solution for sufficiently large $n_{\rm max}$. In such a case, depending on the mass $\M$, we conclude that solutions may feature 
two horizons, one extremal horizon or none. Then, there exists a critical mass $\M_{\rm cr}$ for which the solution describes an extremal black hole of radius $r_{\rm cr}$. This may be obtained by imposing
\eqref{psihp} with $r_0=r_{\rm cr}$ and
\begin{equation}
    f(r_{\rm cr}) = 0 \quad \Longrightarrow \quad \left[\frac{r_{\rm cr}}{r_{\rm S}^{\rm cr}} \right]^{D-3}=\left.\frac{\psi}{h(\psi)} \right|_{\rm cr} \,,
\end{equation}
finding the critical mass $\M_{\rm cr}$ to be 
\begin{equation}\label{eq_critmass}
    \M_\crit =  \frac{1}{2}[h(\psi) r^{D-1}]|_{\crit} \,.
\end{equation}
For $\M>\M_\crit$, the solution is a two-horizon black hole, for $\M=\M_\crit$ we have an extremal black hole whereas for $\M<\M_\crit$ the solution is horizonless and describes a sort of gravitational soliton.

For later use, we observe that the vacuum metric function is an increasing function when
\begin{equation} \label{eq_creciente}
    \frac{\psi h'(\psi)}{h(\psi)} < \frac{D-1}{2},
\end{equation}
which can be verified by differentiating $f(r)$ in \eqref{eq:new_metric}.
%
%
We observe that for Lovelock theories with positive $\talpha_n$ coefficients, there is no critical point due to the fact that
\begin{equation}
\begin{aligned}
    &(D-1)h(\psi) - 2\psi h'(\psi) =\\ & =(D-3)\psi + \sum_{n=2}^{\nmax}(D-1-2n)\talpha_n\psi^{n} >0
\end{aligned}
\end{equation}
iff $\nmax\leq \left\lfloor\frac{D-1}{2}\right\rfloor $.
Namely,~\eqref{psihp} never takes place in that case.
This is a key difference to take into account between Lovelock and QT theories with $n_{\rm max} > \lfloor\frac{D-1}{2}\rfloor$. From this point onward, we will present and discuss results for theories with a critical point. However, the results remain valid for Lovelock theories if we restrict ourselves to the corresponding regime.

As mentioned earlier, a Birkhoff theorem holds in all cases, so the solution outside the star (namely for $r>R$) will be given by the corresponding vacuum solution~\eqref{high_vacuum}. Both metric functions must be continuous at the radius of the star $R$.

\subsubsection{Interior solution}
By plugging~\eqref{eq_perfectfluid} into the field equations~\eqref{eq_QTequations0} and~\eqref{eq_QTequations1}, one finds the following equations for $f(r)$ and $N(r)$,\footnote{For non-constant $h'(\psi)$, it follows that $h'(\psi) = 2\bvrho_{,r}/\psi_{,r}$.}
\begin{align}    \label{eq_h}
h(\psi) & = 2\bvrho \,, \\
\frac{N_{,r}}{r N} & = \frac{(D-1)}{f(r)h'(\psi)}\left(\vrho + \p\right) \,. \label{eq_logNQT}
\end{align}
The first one implicitly determines the function $f(r)$ in terms of the mean density function via the relation $f(r)=1-\psi r^2$.

In order to simplify the computations, we introduce again the Buchdahl variables
\begin{equation}\label{xr2f}
    x \equiv  r^2 \,, \quad \zeta \equiv Nf^{1/2} \,.
\end{equation}
In terms of these, the stress-tensor conservation equation reads
\begin{align}\label{consi1}
\p_{,r} = -\left(\bvrho+\frac{r}{D-1}\bvrho_{,r}+\p\right)\frac{\zeta_{,r}}{\zeta} \,.
\end{align}
%
%
%
Using~\eqref{xr2f}, the equations~\eqref{eq_logNQT} and~\eqref{consi1} can be written as decoupled differential equations for the pressure and the lapse $\zeta$. 
%
%
Eliminating $\zeta$ we obtain the analogous TOV equation for the hydrostatic equilibrium in these theories, whereas eliminating $\p$ we obtain a second order ODE for $\zeta$.
The first reads
\begin{equation}\label{high_TOV_eq}
    \p_{,x} = -\left(\vrho + \p\right) \frac{(D-3)\bvrho\eff+(D-1)\p}{2fh'(\psi)} \,,
\end{equation}
where we defined the ``effective mean mass density'',
\begin{equation}
    (D-3)\bvrho\eff \equiv (D-1)\bvrho-\psi h'(\psi) \,.
\end{equation}
Observe that $\bvrho\eff(r) \geq 0$ as long as $\bvrho(r) \leq \bvrho_\crit \!\equiv\! \M_\crit/r_\crit^{D-1}$, where this critical mean density is obtained when $\psi$ attains the critical value given at~\eqref{psihp}. Moreover, note that for stars with a total positive effective mean mass density, $\bvrho\eff(R) > 0$,~\eqref{eq_creciente} holds, and therefore there is an injective relation between the star's radius $R$ and the blackening factor at the star's surface $f(R)$, as in GR. Going back to the TOV equation, for a given set of couplings $\talpha_n$, one could express $h'(\psi)$ in terms of the radial coordinate, leaving the integrability up to the equation of state---just like in Einstein gravity. On the other hand, making use of the variable $\xi$ defined by $\dif \xi = \dif x/\sqrt{f}$, the equation for the lapse $\zeta$ is
\begin{equation}\label{eq:N_QT}
    \left(h'(\psi)\zeta_{,\xi}\right)_{,\xi} = g(\xi) \zeta \,,
\end{equation}
where we defined a modified version of the function $g(\xi)$ appearing in the Einstein gravity case,\footnote{For non-constant density, it follows that $\bvrho\eff_{,x}=\psi_{,x}\frac{\dif \bvrho\eff}{\dif \psi}$.}
\begin{equation}\label{gxi_eq}
    g(\xi) \equiv \frac{(D-3)}{2}\,\bvrho\eff_{,x}\,.
\end{equation}

 


\subsection{Constant density stars}\label{Sec_QT_Constant_density}
Just as in Einstein gravity, let us consider first the case of stars with constant density
$\vrho=\bvrho=\M/R^{D-1}\equiv\vrho_0$. 
Using the field equation~\eqref{eq_h}, we find that the characteristic polynomial is independent of $r$ in the stellar interior
\begin{equation}
    h(\psi_0) = 2\vrho_0 \equiv h_0 \,.
\end{equation}
Therefore, $\psi_0$ is also constant, and thus the metric function $f(r)$ inside the star is de Sitter-like,\footnote{Having an explicit general expression for $f(r)$ allows us to calculate the proper length along the radial direction in the star's interior, obtaining $\sqrt{\psi_0}\ \rp = \arcsin(\sqrt{\psi_0}\ r).$
As in GR, both variables hold an injective mapping. However, the proper radius $\Rp$ and the areal radius of the star $R$ may not be related by an injective function because $\psi_0$ is a function of $R$.}
\begin{equation}
    f(r) = 1 - \psi_0 r^2 \, , \qquad (r\leq R) .
\end{equation}
In other words, we study the analogous of the Tolman type III solutions in theories with higher-order curvature terms.

Let us write the couplings of the theory as $\talpha_n=\beta_n \alpha^{n-1}$, where $\alpha>0$ is a parameter with units of length$^2$ and $\beta_n$ are some dimensionless couplings. In terms of $\alpha$, one can naturally construct the dimensionless density $2\alpha\vrho_0$ or the (inverse) dimensionless mass $\alpha/\rS^2$, where $\rS=(2\M)^{1/(D-3)}$ is the Schwarzschild radius (of GR). Then, the star will be described in terms of $\psi_0, h_0$ and $h'_0\equiv\frac{\dif h}{\dif \psi}|_{\psi_0}$, which are all independent of $r$ and are functions of the dimensionless quantity $2\alpha\vrho_0$.



Now, define $\Delta$ as the ratio of the effective mean density with respect the mean density
\begin{equation}
\label{eq:defdelta}
    \Delta \equiv\left.\frac{\bvrho\eff}{\bvrho}\right|_{\vrho_0} = \frac{1}{D-3}\left[D-1-2 \frac{\psi_0 h'_0}{h_0}\right] \,,
\end{equation}
which also depends only in $2\alpha\vrho_0$, and equals one for GR. Observe that $\Delta \leq 1$, as $\psi_0 h'_0 \geq h_0$ by the assumed form of $h(\psi)$.



%
%
%
%

%
%


By solving the generalized TOV equation~\eqref{high_TOV_eq} using separation of variables in terms of the variable $r$, one finds the following result for the pressure profile
%
%
\begin{equation} \label{eq:pressureQT}
    \frac{\p(r)}{\vrho_0} = \frac{
    1-\left(\dfrac{f(R)}{f(r)}\right)^{1/2}}
    {\dfrac{(D-1)}{(D-3)\Delta} \left(\dfrac{f(R)}{f(r)}\right)^{1/2} - 1} \,,
\end{equation}
where we imposed the continuity condition on the pressure at the surface of the star, \emph{i.e.}, $\p(R)=0$. Also, the central pressure $\pc$ reads
%
%
%
\begin{equation} \label{eq_pc_QT}
    \frac{\pc}{\vrho_0} = \frac{1 - f(R)^{1/2}}{\dfrac{(D-1)}{(D-3)\Delta} f(R)^{1/2}-1} \,.
\end{equation}
The results would be functionally identical (as functions of the theory-dependent metric functions $f$) to the Einstein gravity ones~\eqref{pressure_GR} and~\eqref{centralp} were not for the presence of $\Delta$ in the denominators.

%
%
%
%
It will be convenient to measure the star radius $R$ in terms of $\rS$. Using~\eqref{eq_pc_QT} we can obtain a formula for the ratio  $R/\rS$ as a function of the central pressure and the star density. It reads
%
\begin{equation}\label{eq_R/rS-pc}
    \left(\!\frac{R}{\rS}\!\right)^{D-3} \!\!=\! \frac{\psi_0}{h_0} \frac{\left(1\!+\!\frac{(D-1)}{(D-3)\Delta}\frac{\pc}{\vrho_0}\right)^2}{\left(1\!+\!\frac{(D-1)}{(D-3)\Delta}\frac{\pc}{\vrho_0}\right)^2\!-\!\left(1\!+\!\frac{\pc}{\vrho_0}\right)^2} \,.
\end{equation}
On the other hand, integrating~\eqref{eq:N_QT} we obtain the expression for $N(r)$ in the star interior,
%
%
\begin{equation}
    N(r) = \dfrac{(D-1) \left(\dfrac{f(R)}{f(r)}\right)^{1/2} - (D-3)\Delta}{(D-1)-(D-3)\Delta} \,,
\end{equation}
%
%
%
which we can use to relate the pressure and $N(r)$, analogously to the Einstein gravity case,
%
\begin{equation}
    \frac{\p(r)}{\vrho_0} = \Delta\frac{(D-3)}{(D-1)}\,\frac{1-N(r)}{N(r)}\, .
\end{equation}
The corresponding alternative expression for the central pressure  follows from setting $r=0$ in the above formula.
Again, these  expressions are functionally identical to the Einstein gravity results~\eqref{eq_GR_pN} up to the appearance of $\Delta$.
%
%
\subsubsection{General aspects of the solution}
Let us examine the sign behavior of the pressure $\p(r)$ and the metric function $N(r)$ in the stellar interior $r\in[0,R)$, as determined by the sign of the $\Delta$ ratio (which depends only on the density $\vrho_0$). 

For $0<\Delta\leq1$, \emph{i.e.,} for positive effective mean density, there are two different stages: a) if $\left(\frac{f(R)}{f(r)}\right)^{1/2}$ is always greater than $\frac{D-3}{D-1}\Delta$ inside the star, then $N(r)$ is positive and the pressure $\p(r)$ remains finite and positive; b) if there exists a radius $r_1$ such that $\left(\frac{f(R)}{f(r_1)}\right)^{1/2}\!=\frac{D-3}{D-1}\Delta$, then the metric function $N(r)$ becomes zero at $r_1$, and the pressure diverges at that point. This case corresponds to a non-acceptable stellar configuration due to the presence of a finite-volume singularity.

On the other hand, at the critical density the ratio vanishes, $\Delta=0$, and therefore $N(r)\geq0$. In the cases where $N(r)$ is positive, the star is pressureless $\p(r)=0$. However, if $N(r) = 0 \Leftrightarrow f(R)=0 \Leftrightarrow R=r_\crit$, \emph{i.e.}, a star with the critical point parameters, we obtain no information about $\p(r)$.

Finally, for negative effective mean densities, $\Delta<0$---that is, for densities larger than the critical one---both functions remain finite in the star's interior: $N(r)>0$ and $\p(r)<0$.

We now analyze the monotonicity of both functions. Seeing that 
\begin{equation}
    \frac{\dif}{\dif r}\frac{f(R)}{f(r)} = \frac{f(R)}{f(r)^2}2\psi_0 r >0 \,,
\end{equation}
in $[0,R)$, we conclude that $N(r)$ is an increasing function in the stellar interior. Similarly, the gradient of the pressure function $\p(r)$ reads
\begin{equation}
    \frac{\dif }{\dif r}\frac{\p(r)}{\vrho_0} = -\Delta\frac{(D-3)}{(D-1)}\frac{N_{,r}}{N^2} \,.
\end{equation}
So, $\p(r)$ is a decreasing function if $\Delta>0$, and an increasing function if $\Delta<0$.

Summarizing: there is an ordinary-matter-stars region with positive pressures, an ``exotic''-matter-stars region with negative pressures, and an unphysical region determined by the stars' singularities at some radius. The boundaries of these regions, along with the non-static region determined by the black hole interior, are:
\begin{enumerate}
    \item \emph{Divergent-central-pressure limit:}  the central pressure diverges when its denominator becomes zero, that is, when $N(0)=0$. The equation that the star parameters must satisfy is
    \begin{equation} \label{eq_buchdahlcondition}
        \sqrt{f(\Rmin)} = \frac{(D-3)}{(D-1)}\Delta \,.
    \end{equation}
     Note that this limit only exists for $\Delta>0$, that is, for densities smaller than $\bvrho_\crit$. Therefore, as mentioned before, it is equivalent to work in terms of $\Rmin$ or $f(\Rmin)$. In order to have an explicit expression of $\Rmin/\rS$ in terms of the density, one can solve $\Rmin$ in the previous expression or take the limit $\pc\to\infty$ in~\eqref{eq_R/rS-pc}. The result reads
    \begin{equation}\label{eq_Rmin/rS}\hspace{2em}
       \frac{\Rmin}{\rS}\!=\!\left[\frac{\psi_0}{h_0} \frac{(D-1)^2}{(D\!-\!1)^2\!-\!(D\!-\!3)^2\Delta^2}\right]^{\frac{1}{D-3}} \,,
    \end{equation}
    or, alternatively, in terms of the gravitational radius,
    %
    \begin{equation}\label{eq_Rmin/r+}\hspace{2em}
         \frac{\Rmin}{\rh} = \left[\frac{(D-1)^2}{(D-1)^2 - (D-3)^2\Delta^2}\right]^{\frac{1}{D-3}} \,.
    \end{equation}
    Interestingly enough, while in GR all constant-density stars had the same compactness limit~\eqref{buchiE}, this is no longer the case for QT theories. As a matter of fact, since $\Delta$ is a monotonically decreasing function of $\vrho_0$---this follows from the structure of $h(\psi)$---, denser limiting stars will strictly be more compact.
    
    \item \emph{Zero-pressure limit:} the pressure vanishes when $\Delta=0$, that is, when $\vrho_0=\bvrho_\crit$.

    \item \emph{Inner-horizon limit:} when the size of the stars coincides with the inner-horizon radius of the would-be black hole,\footnote{It must be the inner horizon, since the outer one is covered by the divergent central-pressure limit.} one finds, using \eqref{eq:pressureQT},
    \begin{equation}
        \lim_{f(R)\to0}\frac{\p}{\vrho_0} = -1 \,.
    \end{equation}
    Therefore the entire pressure profile of the star $\p(r)$ follows a ``Dark Energy''-like equation of state, $\p=-\vrho_0$, without a smooth boundary condition at the star's surface.
    
    \item \emph{Point-like stars limit:} another non-trivial limit corresponds to $R\to0$,  which is the case of arbitrarily small stars. For this, the central pressure vanishes
    \begin{equation}
        \lim_{R\to0}\frac{\pc}{\vrho_0} = 0 \,.
    \end{equation}
    %
\end{enumerate}

We will see instances of all these limits in the explicit examples we study next.

\subsubsection{Energy conditions}
There exist various energy conditions one may consider imposing on the matter stress tensor~\cite{Hawking:1973uf,Poisson_2004,Kontou:2020bta}. Let us first define and comment on each one of them, and after that we will explore the possible existence of excluded regions in the space of constant-density stars.

%
\begin{enumerate}
    \item {\it Weak Energy Condition} (WEC).\\ $T_{ab} v^a v^b \geq 0$ for any future-directed timelike vector $v$. The physical interpretation is clear, namely, the energy density as measured by any observer with $D$-velocity $v^a$ is non-negative. For an isotropic perfect fluid, it becomes:
    \begin{equation}
        \vrho \geq 0 \,, \quad \vrho + \p \geq 0 \,.
    \end{equation}
    Not only the energy density must be non-negative, but the sum of the energy density and pressure must also be non-negative.
    \item {\it Null Energy Condition} (NEC).\\ $T_{ab} k^{a} k^{b} \geq 0$ for any future-directed null vector $k$. This constitutes a limiting case of the WEC condition. 
    For an isotropic perfect fluid, it is evidently less restrictive than the WEC, namely,
    \begin{equation}
        \vrho + \p \geq 0 \,.
    \end{equation}
    Note that the WEC implies the NEC.
  \item {\it Dominant Energy Condition} (DEC).\\
    $T_{ab}v^av^b\geq0$ and $T^{ab}v_a$ is a non-spacelike vector for every timelike vector $v^a$.
    %
    %
    If this condition holds, then the local energy density appears non-negative to any observer and the local energy flow vector is non-spacelike. 
    %
   For an isotropic perfect fluid, it reduces to:
    \begin{equation}
        \vrho \geq |\p| \,.
    \end{equation}
    \item {\it Strong Energy Condition} (SEC). \\ $(T_{ab}-\frac{1}{D-2}g_{ab}T)v^{a}v^{b} \geq0$ for any future directed timelike vector $v^a$. 
    For an isotropic perfect fluid, it reads:
    \begin{equation}\hspace{2em}
        \vrho + \p \geq 0 \,, \quad (D-3)\vrho + (D-1)\p \geq 0 \,.
    \end{equation}
    The form of the second condition is reminiscent of the term which appears in the original TOV equation~\eqref{eq:TOV_GR} for the constant-density case. Since beyond GR this equation gets modified---see~\eqref{high_TOV_eq}---, its motivation in that context is far from clear. Observe that, on general grounds, the SEC may be violated more easily than the WEC~\cite{Kontou:2020bta}. However, the SEC does not imply the WEC.
    %
    %
    %
\end{enumerate}

The relations between the various energy conditions can be schematically summarized as:
$$\rm DEC \Rightarrow WEC \Rightarrow NEC \Leftarrow SEC$$

Making use of the previously derived behavior of $f(r)$ and the form of the pressure profile~\eqref{eq:pressureQT}, in Table~\ref{tab:energy_conditions} we gathered the possible violations of the conditions  depending on the sign of the pressure profiles. While the WEC and the NEC are satisfied for all stars considered, the DEC and the SEC are violated within certain regions in the space of stars. 
\begin{table}[t!]
    \centering
    \begin{tabular}{ c | c c c c}
    \hline
    Type of matter &  DEC & WEC & NEC & SEC \\ 
    \hline
    $\p>0$ & Not for all & \checkmark & \checkmark & \checkmark \\ 
    $\p=0$ & \checkmark & \checkmark & \checkmark & \checkmark  \\ 
    $\p<0$ & \checkmark & \checkmark & \checkmark & Not for all \\ 
    \hline
    \end{tabular}
    \caption{Validity of the energy conditions for constant-density stars depending on the sign of their pressure profiles. A tick means that all stars within the corresponding class satisfy the condition. While  GR fits entirely into the first row as $\Delta=1$ in Einstein gravity, for QT theories $\Delta$ can take negative values and other solutions with different possible violations of the energy conditions appear (last row).
    }
    \label{tab:energy_conditions}
\end{table}
In particular, note that in GR, as well as in any theory in which stars can reach pressures larger than their density, the DEC is not satisfied in those cases. On the other hand, for theories that allow for negative effective densities $(\Delta<0)$, there exist certain stellar configurations that violate the SEC, namely, those for which 
\begin{equation}
    -1 \leq \frac{\pc}{\vrho_0} < - \left( \frac{D-3}{D-1} \right) \,.
\end{equation}
The regions in the space of stars for which the DEC and the SEC are respectively respected have been made explicit in the plots below---see Figs.~\ref{fig_D4} and~\ref{fig_D56}.

\subsubsection{Example 1: Einstein-Gauss-Bonnet gravity}
\begin{figure}[t!]
    \centering
    \includegraphics[width=0.48\textwidth]{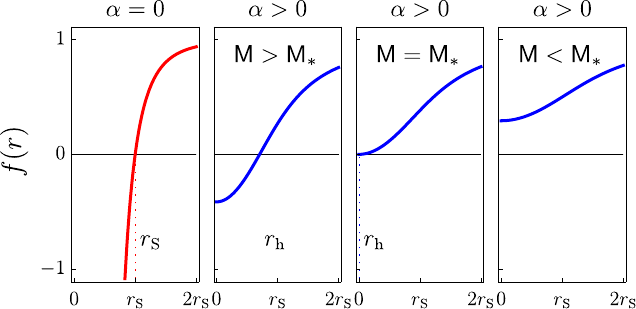}    
    \caption{{We plot the Einstein-Gauss-Bonnet black hole metric function $f(r)$ in $D=5$ for a fixed value of $\alpha$ and different values of the mass. There exists a specific value, $\M_{\ast}=\alpha/2$, (third plot) beyond which smaller values of $\M$ correspond to naked singularities (fourth plot). For greater values of $\M$, on the other hand, the solution describes a black hole with a horizon radius smaller than $r_{\rm S}$ and a curvature singularity milder than Schwarzschild's at $r=0$ (second plot). The red curve in the first plot is the Einstein gravity result.}}
    \label{fig_EGB 5D}
\end{figure}

The simplest possible modification within our framework corresponds to Einstein-Gauss-Bonnet gravity for $D\geq5$. This entails choosing $n_{\rm max}=2$ in the gravitational action. The characteristic polynomial simply reads $h=\psi+\alpha\psi^2$, where we set $\talpha_2=\alpha$. The metric function $f(r)$ in the star's interior and exterior reads, respectively,
\begin{align}
    f(r)&\overset{(r< R)}{=}1-\frac{r^2}{2\alpha}\left(\sqrt{1+8\alpha\vrho_0}-1\right) \, , \\
    f(r)&\overset{(r> R)}{=}1-\frac{r^2}{2\alpha}\left(\sqrt{1+8\alpha \vrho_0\frac{R^{D-1}}{r^{D-1}}}-1\right)\, ,
\end{align}
where recall that $\varrho_0R^{D-1}=\M$.
In $D=5$ there is a specific value of the mass $\M_{\ast}$ beyond which the solution becomes horizonless (although singular at the origin as well)---see Fig.~\ref{fig_EGB 5D}. 
On the other hand, for $D>5$ all vacuum solutions have a horizon~\cite{Boulware:1985wk}.

\begin{figure*}
    \centering
    \includegraphics[width=0.48\textwidth]{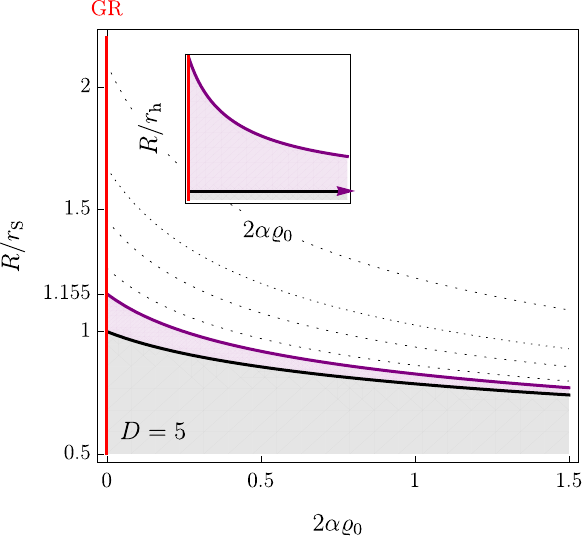}
    \quad
    \includegraphics[width=0.48\textwidth]{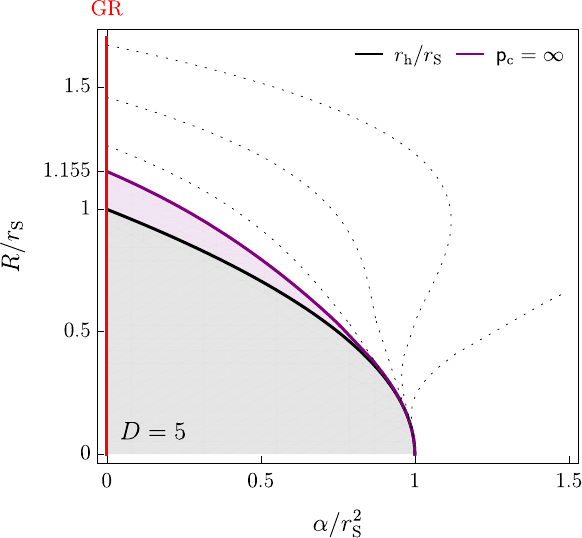}
    \\
    \includegraphics[width=0.48\textwidth]{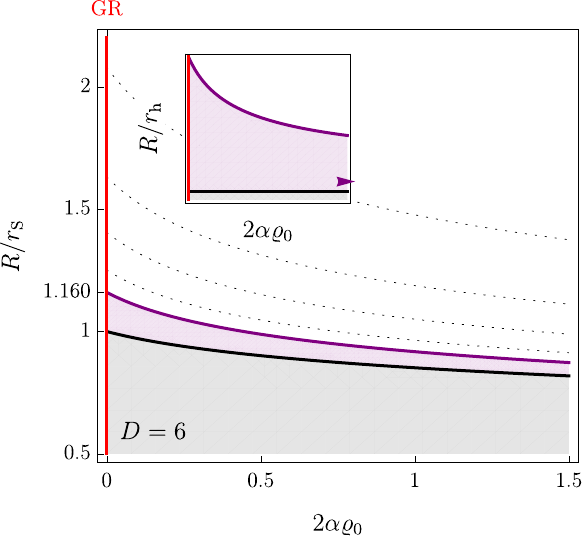}
    \quad
    \includegraphics[width=0.48\textwidth]{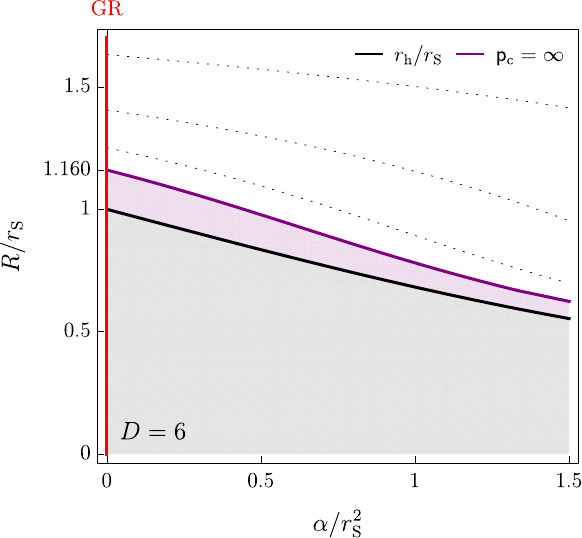}
    \caption{
   Level plots of the central-pressure over the density $\pc/\vrho_0$ (dotted lines) for EGB-like stars in $D=5,6$ with constant density $\vrho_0$. In each of these we can distinguish the following regions: white for stars, purple for singular unphysical stars, and gray for black hole interiors. The boundaries between these regions are represented by thick lines: black for the event horizon and purple for the divergent-central-pressure limit. In all representations, the Y-axis corresponds to the GR limit ($\alpha=0$), whereas the X-axis corresponds to the dimensionless variable $\alpha h(\psi_0)=2\alpha\vrho_0$ in the left plots, and to the inverse of the mass, through $\alpha/\rS^2=\alpha/(2\M)^{2/(D-3)}$ in the right plots. Thus, the vertical lines represent constant densities or constant masses, respectively. As the density grows, the effects of higher-curvature terms become increasingly significant. This is clear in the left plots, where both the gravitational radii and the infinite-central-pressure limits are decreasing functions of the density. The insets show the ratio of the minimal star radius $\Rmin$ to the gravitational radius $r_{\rm h}$ in the same interval; note that the maximum compactness (defined as $r_{\rm h}/\Rmin$) also increases, in both cases, as the density grows. The arrowheads indicates their limiting values as the density approaches infinity:
    for $D=5$, it reaches the maximal degree of compactness, $\Rmin/\rh = 1$,\footnote{Note, however, that it has been demonstrated that in five dimensions there exist no static stars in this limit~\cite{Dadhich:2015rea}.}
    whereas for $D=6$ it is $\Rmin/\rh = (25/24)^{\frac{1}{3}}\approx 1.014$. 
    Regarding the right plots, the interpretation of the lines and regions is analogous: for smaller masses, the maximum compactness increases. It is worth noting that, since the EGB vacuum metric in $D=5$ describes a naked singularity when the mass of the black hole is smaller than $\M_*=\alpha/2$, or, equivalently, when $\alpha/\rS^2 > 1$, the stellar configuration there is no lower bound for the star radius. On the other hand, in higher dimensions, the EGB vacuum metric always describes a black hole. The bottom-right plot, corresponding to six dimensions, shows that the maximum compactness of a star increases as the density grows.
    }
    \label{fig_EGBD5}
\end{figure*}

In Fig.~\ref{fig_EGBD5} we plot in the $R/\rS$--$2\alpha\vrho_0$ plane curves of constant central-pressure and the relevant limits. The $\Delta$ ratio is in this case given by
\begin{equation}
    \Delta = 1-\frac{2}{D-3}\left(1-\frac{\sqrt{1+8\alpha\vrho_0}-1}{4\alpha\vrho_0}\right) \,.
\end{equation}
%
Interestingly, the maximum compactness limit for large densities is rather similar to the Einstein gravity result, it reads
\begin{equation}
    \lim_{\vrho_0\to\infty}\left(\frac{\Rmin}{\rh}\right)^{D-3} = \frac{(D-1)^2}{8(D-3)} \,,
\end{equation}
%
but note that these stars are always more compact than in GR.


%
%

\subsubsection{Example 2: High-density limit of Lovelock gravities}
\begin{figure}[t!]
    \centering
    \includegraphics[width=0.48\textwidth]{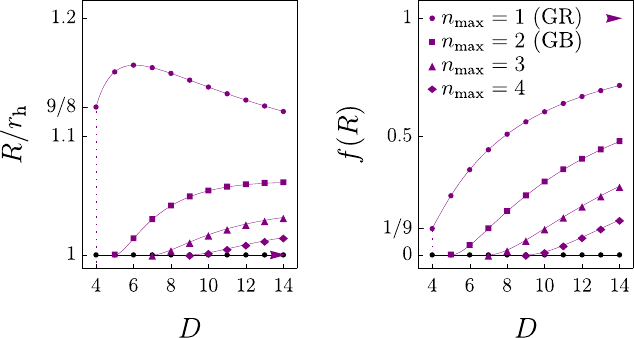}
    \caption{ {Maximum compactness limit for constant-density stars in single Lovelock gravities as as a function of the spacetime dimension $D$. We follow the color convention of Fig~\ref{fig_GR_Buchdahl}. In both plots, we include again the general relativity limit, which corresponds to the case $\nmax=1$. Furthermore, we have depicted the limiting cases for the three first orders of single Lovelock gravities, corresponding to $\nmax=2,3,4$, where the second order is Gauss-Bonnet. In both plots, the arrowheads indicate this limit for all the cases as the dimension approaches infinity. Although Lovelock theories can be built with $\nmax\leq\lfloor\frac{D-1}{2}\rfloor$, it has been shown that there are no static stars in odd dimensions when $D=2\nmax+1$~\cite{Dadhich:2015rea}. Therefore, the minimal dimension that allows stable stars in a single Lovelock theory of order $\nmax$ is $D=3\nmax+1$.
    }}
    \label{fig_Lovelock_Buchdahl}
\end{figure}

Consider now a Lovelock theory including invariants  up to order $\nmax=\lfloor\frac{D-1}{2}\rfloor$---see Fig.~\ref{fig_Lovelock_Buchdahl}. 
Due to the fact that $h(\psi)$ is an increasing function of $\psi$, taking the $\vrho_0\to\infty$ limit is equivalent to taking $\psi_0\to\infty$ in the constant density case. Thus
\begin{equation}
    \lim_{\psi_0\to\infty}\Delta = \frac{D-1-2\nmax}{D-3} \,,
\end{equation}
and then,
\begin{equation}
    \lim_{\vrho_0\to\infty}\left(\frac{\Rmin}{r_{\rm h}}\right)^{D-3} = \frac{(D-1)^2}{4\nmax(D-1-\nmax)} \,.
\end{equation}
Observe that this limit is equivalent to studying theories consisting solely of a unique Lovelock density of order $\nmax$~\cite{Dadhich:2016fku}.


\subsubsection{Example 3: Hayward stars in QT gravities}
In this subsection we consider classes of QT theories that admit the $D$-dimensional Hayward black hole~\cite{Hayward:2005gi} as their unique spherically symmetric vacuum solution, as well as others for which the only solution is  a generalized version of such metric~\cite{Bueno:2024zsx}. The characteristic polynomial  reads
%
\begin{equation}
    h_{\N}(\psi) = \frac{\psi}{\left[1-(\alpha\psi)^\N\right]^{1/\N}} \,,
    \label{Haywardth}
\end{equation}
%
where $\alpha$ sets the length-scale of the theory and $\M$ the ADM mass. 
From this it is straightforward to find the explicit form of all the action couplings in terms of $\alpha$. The metric function reads in turn
\begin{equation}
    f_\N(r) = 1 - \frac{2\M r^2}{\left[r^{\N(D-1)}+(2\M\alpha)^{\N}\right]^{1/\N}} \,,
\end{equation}
which reduces to the usual Hayward metric for $\N=1$. It is worth noting that in even dimensions, the lowest possible value of $\N$ is $D/2$~\cite{Bueno:2024zsx}. 

\begin{figure}[t!]
    \centering
    \includegraphics[width=0.48\textwidth]{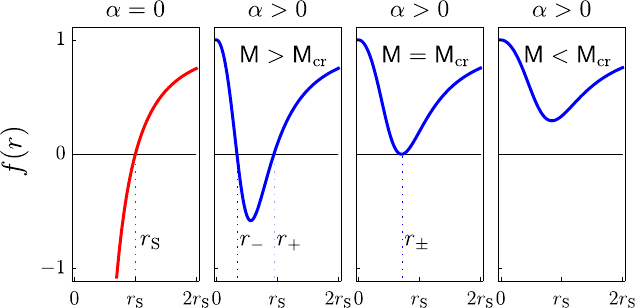}
    \caption{{We plot the metric function $f(r)$ of the five-dimensional Hayward spacetime for different values of the mass. In the first plot (red curve), the coupling constant is set to zero, representing the GR solution of the singular black hole (Schwarzschild-Tangherlini solution). In the subsequent three plots (blue curves), we show the regular solution for different values of the total mass: a) when $\M>\M_\crit$, the solution has both an outer and an inner horizon. Note that the gravitational radius of this black hole is smaller than $\rS$; b) when $\M=\M_\crit$, both horizons degenerate, which corresponds to the extremal Hayward black hole, with mass $\M_\crit$ and gravitational radius $r_\crit$; c) when $\M<\M_\crit$, the solution is horizonless.}}
    \label{fig_Hayward}
\end{figure}

These vacuum solutions correspond to regular black holes with two horizons located at $r_\pm$ for $\M>\M_\crit$, where the critical mass is given by
\begin{equation}
    \M_\crit = \frac{(D-1)^{\frac{1}{\N}}}{2^{\frac{\N+1}{\N}}}\left(\frac{D-1}{D-3}\right)^{\frac{D-3}{2\N}}\alpha^{\frac{D-3}{2}} \,.
\end{equation}
For the critical mass the two horizons coincide at
\begin{equation}
    \frac{r_\crit}{\rS^\crit} = \left(\frac{2}{D-1}\right)^{\frac{1}{\N(D-3)}} \,,
\end{equation}
resulting in a degenerate horizon. We define the critical mean density $\bvrho_\crit$ as follows
\begin{equation}
    \bvrho_\crit = \frac{\M_\crit}{r_\crit^{D-1}} = \frac{1}{2\alpha}\left(\frac{D-3}{2}\right)^{1/\N} \,.
\end{equation}
And the last case, for $\M<\M_\crit$, the solution is horizonless. These three possibilities are shown in Fig.~\ref{fig_Hayward}.

For Hayward-like stars with constant density, the stellar interiors we consider are described by the following de Sitter-type radial metric functions in odd and even $D$, respectively,
\begin{align}
    f^{\rm odd}(r) & = 1 - \frac{2\M r^2}{R^{D-1}+2\M\alpha} \,,\\ \notag
    f^{\rm even}(r) & = 1 - \frac{2\M r^2}{\left[R^{D(D-1)/2}+(2\M\alpha)^{D/2}\right]^{2/D}} \,.
\end{align}
Namely, in odd dimensions we simply consider the $\N=1$ solution, whereas in even dimensions we choose $\N=D/2$.

Another relevant quantity to keep in mind is the effective density ratio, which in this case reads
\begin{equation}
    \Delta = 1 - \frac{2}{D-3}(2\alpha\vrho_0)^{\N} \,.
\end{equation}

\begin{figure*}
    \centering
    \includegraphics[width=0.48\textwidth]{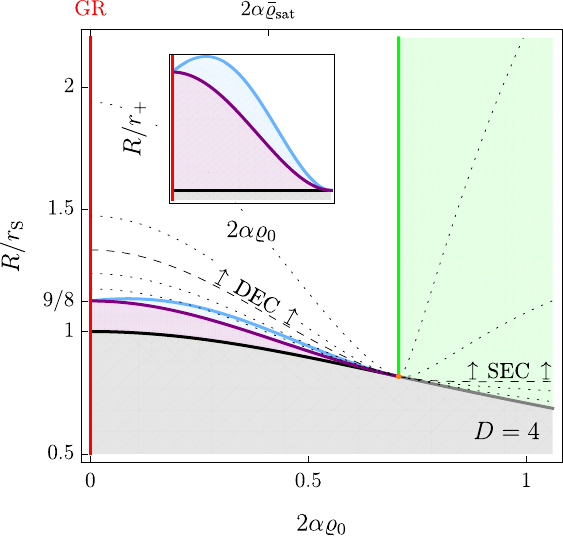}
    \quad
    \includegraphics[width=0.48\textwidth]{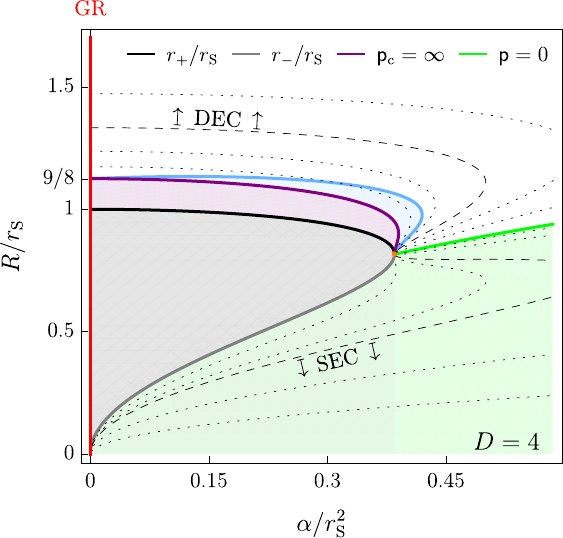}
    \caption{
    Level plots of the central-pressure over the density $\pc/\vrho_0$ (dotted lines) for Hayward-like stars with constant density $\vrho_0$ in $D=4$. In each of these we can distinguish the following regions: white for ordinary matter stars, green for exotic matter stars, purple for singular unphysical stars, and gray for black hole interiors.\footnote{The meaning of the light blue curves and regions will be explained in Sec.~\ref{sec_5}. For now, both can be though of as belonging to the white region, corresponding to ordinary matter stars. Also,  $\bvrho_{\rm sat}$ is defined and explained in subsection~\ref{buchineq}.} The boundaries between these regions are represented by thick lines: black for the outer horizon, gray for the inner horizon, purple for the divergent-central-pressure limit and green for the zero-pressure limit. And the intersection of all of them occur at the critical point (orange dot). In addition, we have depicted with dashed lines the constant central-pressure lines that saturate the DEC and SEC; those energy conditions are respectively satisfied in the regions indicated with the arrows. In all representations, the Y-axis corresponds to the GR limit ($\alpha=0$), whereas the X-axis corresponds to the dimensionless variable $\alpha h(\psi_0)=2\alpha\vrho_0$ in the left plot, and to the inverse of the mass, through $\alpha/\rS^2=\alpha/(2\M)^2$ in the right plot. Thus, the vertical lines represent constant densities or constant masses, respectively. As the density grows, the effects of higher-curvature terms become increasingly significant. This is clear in the left plot, where both the gravitational radii and the infinite-central-pressure limits are decreasing functions of the density, up to the critical one. In fact, note that the maximum compactness (defined as $r_+/\Rmin$) also increases as the density approaches the critical value, resulting in fully compact ordinary matter stars near critical values. For even higher densities, constant density stars must have negative pressures. An interesting feature which can be distinguished in the right plot is that the minimum mass for which infinite pressure stars exist does not coincide with $\M_{\rm cr}$, but it takes a different non-analytical value which we denote by $\M_{\rm mm}, R_{\rm mm}$. This corresponds to the tip of the purple curve (which becomes more noticeable as $D$ and/or $\N$ increases). For masses greater than $\M_{\rm mm}$ and smaller than $\M_{\rm cr}$ (\emph{i.e.,} moving towards the left in the plot, remember that $\rS=2\M$) there exist two radii with infinite-central-pressure (as the radius is reduced, one enters the purple region and then leaves it again into a region of ordinary matter stars, before reaching the exotic matter stars regime). For $\M>\M_{\rm cr}$, the exotic stars are covered by two horizons, making them completely indistinguishable from a black hole to an external observer (this region has been shaded in a lighter green). The relevant quantities for $D=4$, $\N=2$, Hayward-like stars read: $2\alpha\bvrho_\crit=1/\sqrt{2}\approx 0.707$, $r_\crit/\rS^\crit=(2/3)^{\frac{1}{2}}\approx 0.816$, $\alpha/(\rS^\crit)^2=(2^2/3^3)^{\frac{1}{2}}\approx 0.385$, $R_{\rm mm}/\rS^{\rm mm}\approx 0.900$, $\alpha/(\rS^{\rm mm})^2\approx0.390$, and $2\alpha\bvrho_{\rm sat}=1/\sqrt{6}\approx 0.408$.
    }
    \label{fig_D4}
\end{figure*}

\begin{figure*}
    \centering
    \includegraphics[width=0.48\textwidth]{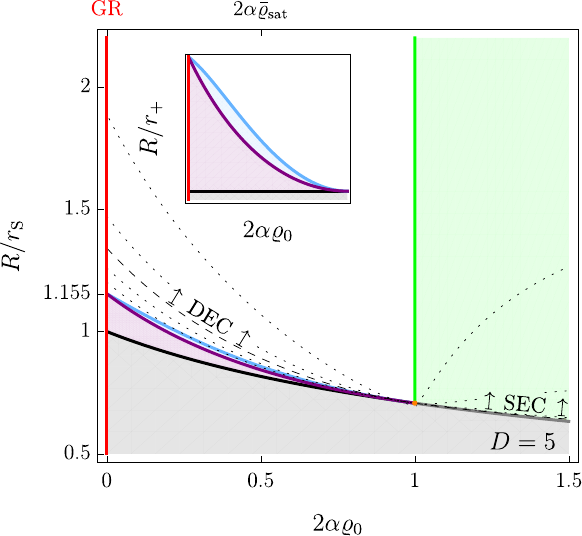}
    \quad
    \includegraphics[width=0.48\textwidth]{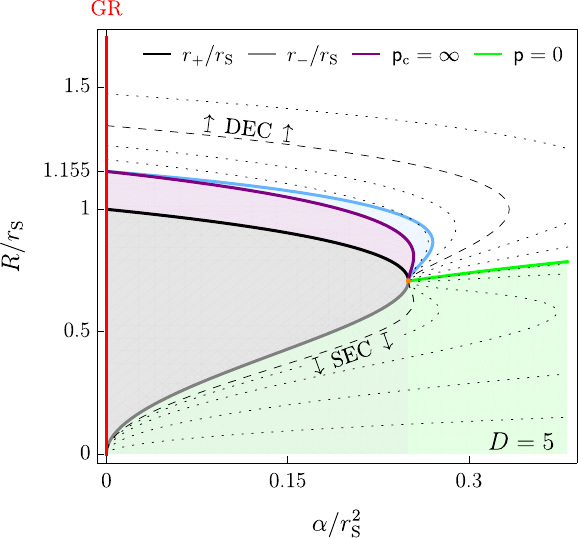}
    \\
    \includegraphics[width=0.48\textwidth]{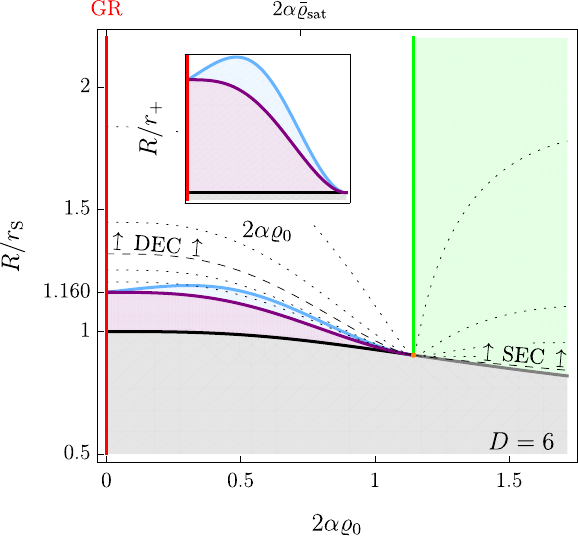}
    \quad
    \includegraphics[width=0.48\textwidth]{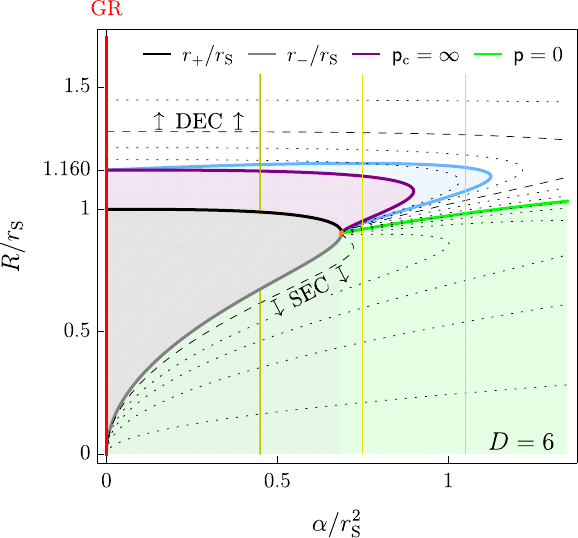}
    \caption{
   { Level plots of the central-pressure over the density $\pc/\vrho_0$ for Hayward-like stars with constant density $\vrho_0$ in $D=5$ (top) and $D=6$ (bottom). We use the same color convention as in Fig.~\ref{fig_D4}. Each yellow vertical line in the right-bottom plot represents a constant-mass star solution. The relevant quantities for the $D=5$, $\N=1$, Hayward-like stars read: $2\alpha\bvrho_\crit=1$, $r_\crit/\rS^\crit=1/\sqrt{2}\approx0.707$, $\alpha/(\rS^\crit)^2=1/4$, $R_{\rm mm}/\rS^{\rm mm}\approx0.807$, $\alpha/(\rS^{\rm mm})^2\approx0.254$, and $2\alpha\bvrho_{\rm sat}=1/2$.
    Similarly, for $D=6$, $\N=3$, Hayward-like stars, the relevant quantities read:
    $2\alpha\bvrho_\crit=(3/2)^{\frac{1}{3}}\approx 1.145$, $r_\crit/\rS^\crit=(2/5)^{\frac{1}{9}}\approx 0.903$, $\alpha/(\rS^\crit)^2=(2^2 3^3/5^5)^{\frac{1}{9}}\approx 0.688$, $R_{\rm mm}/\rS^{\rm mm}\approx1.074$, $\alpha/(\rS^{\rm mm})^2\approx0.899$, and $2\alpha\bvrho_{\rm sat}=3^{\frac{1}{3}}/2\approx 0.721$.
    }}
    \label{fig_D56}
\end{figure*}

Using these functions and the previous results, in Figs.~\ref{fig_D4} and~\ref{fig_D56} 
we present two contour plots of the central pressures for the $D=4,5,6$ cases. Within each of those plots we can distinguish four regions corresponding, respectively, to: ordinary matter stars (white and light blue regions);  negative-central pressure stars (green region);  unphysical stars for which curvature singularities would occur in the interior (purple region); black hole interiors (gray region). Detailed comments on the structure of this space of solutions can be found in the captions. 

It is worth recalling that the maximum compactness limit, corresponding to a divergent-central-pressure, only exists for positive effective densities. In fact, as the density approaches the critical value, the maximum compactness tends to reach the extremal black hole size. On the other hand, no infinite pressure  limit exists for stars with greater densities (\emph{i.e.,} with negative effective densities). In the case of negative-effective-density stars with masses greater than the critical value, the solutions are covered by two horizons, so the solutions represent a black hole from the point of view of an exterior observer. Hence, in that case the stars act as the exotic-matter core of a black hole.

\begin{figure*}
    \centering
    \includegraphics[width=0.96\textwidth]{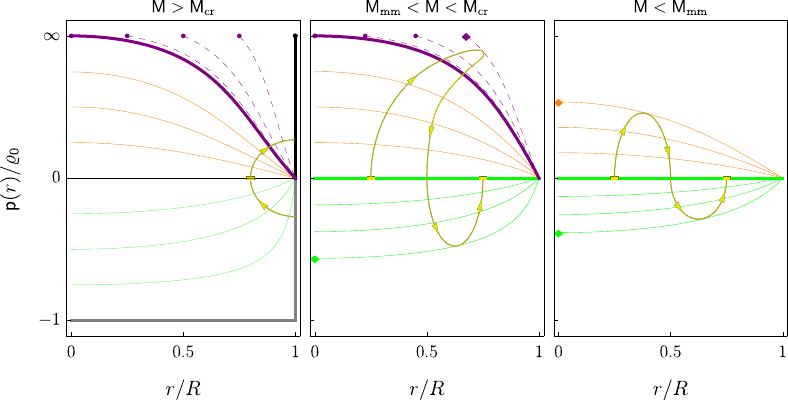}
    \caption{
    {In the case of constant-density Hayward stars in $D=6$, we plot three families of pressure profiles for star masses within the three different regions defined by $\M_\crit$ and $\M_{\rm mm}$---see the yellow lines in Fig.~\ref{fig_D56}. 
    First, the yellow arrows indicate the direction in which the star's radius $R$ decreases in each case, starting from a pressureless star with a very large radius all the way to a point-like star (avoiding the dynamical configurations due to the black hole interior). From the level curves of the central-pressure in Figs.\,\ref{fig_D4} and~\ref{fig_D56}
    , one realizes that there may be two possible radii for the same pressure profile. Orange lines correspond to ordinary matter stars, purple dashed lines to unphysical ones and green ones to exotic matter stars. The limit of divergent central-pressure corresponds to the solid purple line. This presents a point-like singularity at $r=0$ (indicated with a dot). On the other hand, the purple dashed lines are unphysical stars which would possess finite-volume singularities. The limiting case is depicted with a solid black line, which corresponds to the configuration with $R=r_+$, presenting a singularity at the surface of the unphysical would-be star.  The inner-horizon limit is represented by a solid gray line and the pressureless limit is depicted with a solid green line.
    (Left) Stars with $\M>\M_\crit$. This plot is an obvious generalization of the Einstein gravity case---see Fig.~\ref{fig_profilesGR}. We observe the existence of exotic matter configurations involving negative pressures for stellar radii smaller than the inner horizon radius. (Center) Stars with $\M_{\rm mm}<\M<\M_\crit$. The interpretation is analogous to the one of the left plot, with a big main difference: here there is no black hole region and two bounds appear---an upper one, corresponding to the maximum volume singularity (purple diamond), and a lower bound, corresponding to the minimum central-pressure (green diamond). Furthermore, the yellow arrow crosses twice each profile, meaning that there are two different radii with the same ratio $\p(r)/\vrho_0$. The pressureless star case can be achieved both with a very large radius, with a zero-size radius, and with a finite radius corresponding to the green line limit. (Right)  Stars with $\M<\M_{\rm mm}$. The analysis is analogous, but now there is no longer a bound associated with an infinite-central-pressure. The upper bound corresponds now to a specific central pressure, indicated with an orange diamond.
    }}
    \label{fig_profilesQT}
\end{figure*}

Additionally, in Fig.~\ref{fig_profilesQT}  we have presented the pressure profiles $\p(r)$ for stars belonging to the various regions in the space of parameters. Analogously to the Einstein gravity case displayed in Fig.~\ref{fig_profilesGR}, we have compactified the positive part of the Y-axis using an $\arctan$ function. Hence, when the pressure reaches the top of the diagram, it means that such star would involve a singularity at the radial coordinate point marked with a dot. Detailed comments on the profiles appear in the captions.


\clearpage

\subsection{Buchdahl's inequalities}\label{buchineq}
In this subsection we study Buchdahl's limit for QT gravities in the case of stars whose mean mass-density decreases outwards $\bvrho_{,x}\leq0$ (which implies $\psi_{,x}\leq0$ because $h'(\psi)\geq0$).

In Buchdahl's original work, such assumption implies that $g(\xi)$ is non-positive. In the present case, on the other hand, $g(\xi)$ will be negative if and only if $\dif \bvrho\eff/\dif \psi$
is positive. Using the definition of the characteristic polynomial, this derivative can be expressed as
\begin{equation}
\begin{aligned} \label{eq_gxifunctionQT}
    &2(D-3)\frac{\dif\bvrho\eff}{\dif \psi} = (D-3)h'-2\psi h'' \\ 
    & = (D-3)+\sum_{n=2}^{\nmax}(D-1-2n)n\talpha_n\psi^{n-1}\,.
\end{aligned}
\end{equation}
In the following, we will also be assuming that $\dif\bvrho\eff/\dif \psi \geq 0$. While this is true in full generality for Lovelock gravities---as these theories do not exist for curvature orders above $\nmax\leq\lfloor\frac{D-1}{2}\rfloor$---, the monotonicity of $h(\psi)$ in terms of $\psi$ will only be guaranteed for a restricted set of mean density profiles in generic QT theories. 
Indeed, the procedure remains valid for mean densities smaller than a specific value of saturation $\bvrho_{\rm sat}$, determined by the chosen theory. Due to the monotonicity of the characteristic polynomial, the saturation density is unique and satisfies: $0 < \bvrho_{\rm sat} < \bvrho_\crit$.\footnote{Notice that in Lovelock theories there is neither a critical density nor a saturation density.} For instance, for Hayward stars of any order in any dimensions this value reads:
\begin{equation} \label{eq:density_sat}
    \bvrho_{\rm sat}\!=\!\frac{1}{2\alpha}\left(\frac{D-3}{2(\N+1)}\right)^{1/\N}\!\!=\!\frac{\bvrho_\crit}{(\N\!+\!1)^{1/\N}}\!<\!\bvrho_\crit \,.
\end{equation}
%


Let us proceed to obtain a bound for the maximum compactness for static perfect-fluid stars in QT gravities. As previously mentioned, for total densities smaller than the critical one, there is a equivalence between the star radius $R$ and its blackening factor at the surface $f(R)$. Therefore, one can use either $R$ or $f(R)$ to discuss Buchdahl limits. Now, we follow Buchdahl's steps to find a lower bound for $f(R)$. We note that 
\begin{equation} \label{eq:QT_inequality1}
    (h'\zeta_{,\xi})_\ce \geq h'\zeta_{,\xi} \geq (h'\zeta_{,\xi})_\bo \,,
\end{equation}
where the subscripts ``c'' and ``b'' represent the evaluation at the center and at the boundary of the star, respectively. This relation follows from the fact that the RHS of~\eqref{eq:N_QT} is negative. We can determine $(h'\zeta_{,\xi})_\bo$ by using the metric functions in the vacuum, as we are at $r=R$:
\begin{equation}
    (h'\zeta_{,\xi})_\bo = \left.\frac{\dif h}{\dif \psi}\frac{\dif \sqrt{f}}{\dif x/\sqrt{f}}\right|_{r=R} = \frac{D-3}{2}\bvrho_\bo\eff \,.
\end{equation}
Therefore, from the second inequality in~\eqref{eq:QT_inequality1}, one finds
\begin{equation} \label{eq:QT_inequality2}
    h'(\psi)\zeta_{,\xi}\dif \xi  \geq \frac{D-3}{2}\bvrho_\bo\eff \dif \xi \,.
\end{equation}
We now integrate both sides, finding an upper bound for the LHS and a lower bound for the RHS.

First, under Buchdahl's assumption, it follows that $h'(\psi)_{,x} \leq 0$, and therefore $h'(\psi_\ce) \geq h'(\psi)$. Hence, the LHS is bounded by
\begin{equation}
    \int_\ce^\bo h'(\psi)\zeta_{,\xi} \dif\xi \leq h'(\psi_\ce) (\zeta_\bo-\zeta_\ce) \,.
\end{equation}
Whereas on the RHS, making use again of Buchdahl's assumption, it is straightforward that $\sqrt{1-\psi r^2}\leq \sqrt{1-\psi_\bo r^2}$ holds for all $r\in[0,R)$. Consequently, it can bounded as
\begin{equation}
    \bvrho\eff_\bo\int_\ce^\bo \frac{r\dif r}{\sqrt{f}} \geq\frac{\bvrho\eff_\bo}{\psi_\bo}\left(1-\sqrt{f(R)}\right)\,.
\end{equation}
Putting it all together, the inequality~\eqref{eq:QT_inequality2} becomes
\begin{equation}
    \zeta_\bo - \zeta_\ce \geq (D-3) \frac{\bvrho\eff_\bo}{\psi_\bo h'_\ce} \left(1-\sqrt{f(R)}\right) \,,
\end{equation}
where the equality holds in the constant density case due to the saturation of Buchdahl's assumption.

Rewriting it in terms of $f(R)$ and substituting $f(0)=1$ and $N(R)=1$, the inequality reads
\begin{equation}
    f(R) \geq \left( \frac{(D-3)\bvrho\eff_\bo + \psi_\bo h'_\ce N(0)}{(D-1)\bvrho_\bo+\psi_\bo(h'_\ce-h'_\bo)} \right)^2 \,.
\end{equation}
The maximum compactness for a specific density profile is reached when the value of $N(0)$ is minimized, which occurs at certain $\Rmin$ when the metric is singular at the center. Therefore, $f(\Rmin)$ of the most compact stars in a high-curvature theory is bounded by
\begin{equation}
\label{boundQT}
    f(\Rmin) \geq  \frac{\left(\dfrac{D-3}{D-1}\dfrac{\bvrho\eff_\bo}{\bvrho_\bo}\right)^2}{\left(1+\dfrac{\psi_\bo(h'_\ce-h'_\bo)}{(D-1)\bvrho_\bo}\right)^2} \,,
\end{equation}
where the equality holds for the case of constant density stars---for which $h'(\psi_\ce)=h'(\psi_\bo)$, so that the denominator becomes one and we eventually recover expression~\eqref{eq_buchdahlcondition} obtained in Sec.~\ref{Sec_QT_Constant_density}. For strictly decreasing density-profiles, the expression becomes a strict inequality. Note that the bound depends on the density at the surface and at the center of the star.\footnote{A similar dependence had previously been observed in EGB gravity~\cite{Wright:2015yda}. However, it is worth noting that five-dimensional EGB is the only theory with higher-curvature terms in which the RHS of Buchdahl's inequality does not depend on the total density $\bvrho_\bo$.}

%

From~\eqref{boundQT}, it is not clear which density profile gives the minimum value for the bound of $f(\Rmin)$ among all possible density profiles that are outward-decreasing and are below the saturation density $\bvrho_{\rm sat}$. The reason for this is the following: since the RHS of the inequality is density-dependent, it may well happen that a non-constant density configuration yields a lower bound for $f(\Rmin)$ which is unattainable for any density profiles\footnote{For instance, suppose that for a certain total density $\bvrho_\bo$ we obtain the following bounds: $f(\Rmin)=0.2$ for the constant-density profile, and $f(\Rmin)>0.1$ for some decreasing profile with higher central density. However, it turns out that no decreasing profile density reaches a blackening factor below $0.2$. Hence, the constant-density configuration is more compact, even though the Buchdahl bound is smaller (and unreachable; note the strict inequality sign) for decreasing densities.}---which are the ones that saturate the inequality~\eqref{boundQT}. As a result, we cannot exclude from these arguments the existence of theories for which certain non-constant profile densities provide lower values for $f(\Rmin)$ than any other constant density configurations.

A better bound with no explicit dependence on the value of the central density may be obtained when the saturation density exists.\footnote{Note that, for Lovelock theories, we cannot proceed in the same way because there is no maximum value for the density. In fact, as shown for constant density configurations, in some cases $f(\Rmin)$ can reach zero.} Noticing that $h'(\psi)$ is an increasing function of $\psi$, its maximum value will be attained when the density acquires its maximum value: the saturation density $\bvrho_{\rm sat}$.
Therefore, stars whose central density equals the saturation one will present a higher compactness, so $f(\Rmin) \geq  f(\Rmin)|_{h'_{\rm c}=h'_{\rm sat}}$. Buchdahl's inequality then takes the form
\begin{equation}\label{boundQTsat}
    f(\Rmin) \geq \frac{\left(\dfrac{D-3}{D-1}\dfrac{\bvrho\eff_\bo}{\bvrho_\bo}\right)^2}{\left(1+\dfrac{\psi_\bo(h'_{\rm sat}-h'_\bo)}{(D-1)\bvrho_\bo}\right)^2} \,,
\end{equation}
where the equality is attained in the constant-density case of $\bvrho=\bvrho_{\rm sat}$. In the last inequality, Buchdahl's limit is understood as the maximum compactness reached for a given total density $\bvrho_{\rm b}$.

Note that the RHS of~\eqref{boundQTsat} cannot be arbitrarily small. Since $\Delta \equiv\frac{\bvrho\eff_\bo}{\bvrho_\bo}$ is such that $1 > \Delta \geq \Delta_{\rm sat}$ and $\frac{\psi_\bo}{\bvrho_\bo} < 2$, we obtain
\begin{equation}\label{boundQTnminsat}
    f(\Rmin) >  \frac{(D-3)^2 \Delta_{\rm sat}^2}{\left(D-3+2 h'_{\rm sat}\right)^2} >0\,.
\end{equation}
This lower bound is unattainable, but it proves that $f(\Rmin)$ cannot be made arbitrarily small if the density is always below its saturation value. 

\subsubsection{Buchdahl limits}
One can go even further and find a stricter (and attainable) Buchdahl limit in some cases; that is, determine the most compact configuration (within Buchdahl's assumptions), and see whether there exists a specific set of star parameters and pressure profile such that $f(\Rmin)$ is a global minimum. Our strategy is as follows. Observe first that the RHS is determined by the theory---through $h(\psi)$---and it depends on the total density $\bvrho_{\rm b}$. Then, in the case of theories for which the RHS is a decreasing function of $\bvrho_\bo$, the minimum bound will correspond to stars with $\bvrho_\bo = \bvrho_{\rm sat}$, that is, to stars with constant-density equal to the saturation value. This constitutes a sufficient condition to prove that the corresponding constant-density configuration represents the highest degree of compactness. As a result, we conclude that stars in this restricted class of QTs  will have a maximum compactness attained by constant-density profiles equal to the saturation value. This is to be compared to the GR result, where the corresponding limiting maximum compactness is achieved for \emph{any} constant-density Buchdahl star.

 In this restricted set of theories, any star with mass $\M$, radius $R$, finite pressure and non-increasing mean density (which does not exceed the saturation value in any interior point $r$, $\bvrho(r)\leq \bvrho_{\rm sat}$) must satisfy
\begin{equation}
    f(R) > f(\Rmin) \geq f(R_{\rm Buch.}) \,,
\end{equation}
where $f(\Rmin)$ is the blackening factor of the limiting configuration for this specific mass $\M$ and density profile, and $f(R_{\rm Buch.})$ is the value of the Buchdahl limit, corresponding to the saturation constant-density star:
\begin{equation}
    f(R_{\rm Buch.}) = \left(\frac{D-3}{D-1}\right)^2 \Delta_{\rm sat}^2 \,,
\end{equation}
which is always smaller than in GR. 

As an explicit example, let us once again consider QT theories whose solutions are given by Hayward-like stars. In this case, the RHS of~\eqref{boundQTsat} is a decreasing function provided that $\N\geq \lfloor\frac{D-3}{2}\rfloor$. In this subclass of models, the maximum compactness reached by any star, under Buchdahl's assumptions, is bounded by a constant-density star with saturation density. Its corresponding blackening factor at the surface reads
\begin{equation}
    f(R_{\rm Buch.}) = \left(\frac{D-3}{D-1}\right)^2 \left(\frac{\N}{\N+1}\right)^2 \,.
\end{equation}
%
Equivalently, using the result~\eqref{eq_Rmin/r+} for constant-density stars, one obtains the following expression for the radius of the global Buchdahl limit
\begin{equation}
    \frac{R_{\rm Buch.}}{r_+^{\rm sat}} = \frac{1}{\left(1-\dfrac{(D-3)^2\N^2}{(D-1)^2(\N+1)^2}\right)^{1/(D-3)}} \,,
\end{equation}
where $r_+^{\rm sat}$ is the gravitational radius of the event horizon of the Hayward black hole for mass equal to $\bvrho_{\rm sat} R_{\rm Buch.}^{D-1}$.

\subsection{Markov's limiting curvature hypothesis} \label{sec_5}

In~\cite{PismaZhETF.36.214}, Markov proposed that there should exist an upper bound on the maximum curvature attainable in the universe. According to this hypothesis, all curvature invariants should always remain below certain universal values determined by some fundamental scale $\ell$.\footnote{In Markov's original proposal, $\ell$ was identified with Planck's length.} For example, the Kretschmann scalar would satisfy
%
\begin{equation}\label{Km}
    K=R_{abcd}R^{abcd}\leq1/\ell^4 \,.
\end{equation}
Naturally, a proposal of this type is very suggestive in the context of regular black holes, since in that case all curvature invariants remain bounded for every solution.  However, the fact that a single universal bound such as~\eqref{Km} may hold for every regular black hole of a given theory regardless of its mass is  highly non-trivial---for instance, the Bardeen solution~\cite{1968qtr..conf...87B} fails to satisfy this condition. In the context of QT gravities, 
Frolov, Koek, Pinedo Soto, and Zelnikov proved that the limiting curvature hypothesis holds for general QT theories admitting regular black holes~\cite{Frolov:2024hhe}---see also~\cite{Bueno:2024zsx}. Their argument relies on the observation that any curvature invariant within these theories satisfies a certain general  \emph{scaling property} in the vacuum. This states that all the curvature invariants depend on a single dimensionless quantity, $2\M\alpha/r^{D-1}$. As an example, the Kretschmann scalar of the Hayward black hole (for $\N=1$) is uniformly bounded as
\be 
K \le K_{\rm max} = \frac{2D(D-1)}{\alpha^2} \, .
\ee

In a companion paper~\cite{QTvaccine}, we shall study in more detail the boundedness of curvature invariants of static solutions in Quasi-topological gravity minimally coupled to matter. Here, we shall simply examine the Kretschmann scalar for the configurations studied so far. Before doing so, let us make a few qualifying remarks. First of all, let us note that in the presence of minimally coupled matter, the limiting curvature hypothesis no longer holds in the following sense: the curvature invariants, if bounded, are no longer bounded in a solution-independent manner. In other words, the inclusion of minimally coupled matter spoils the universal upper bound on the Kretschmann scalar that holds in vacuum.

The second comment we make concerns the generality of the minimally coupled matter assumption. On pragmatic grounds, this has been useful to make explicit calculations possible. However, it is also unnatural in the following sense. In the gravitational sector we have considered an infinite tower of higher-curvature corrections sufficiently general to describe the vacuum gravitational effective field theory. When coupling such a theory to matter, the most natural thing to do would be to consider a full tower of corrections in the matter sector as well. However, this is rather difficult as aside from certain special cases~\cite{Cano:2020qhy, Cano:2020ezi, Fernandes:2025fnz, Fernandes:2025mic}, there are very few known examples of full gravity + matter resummations with controllable dynamics. Of those constructions where this is possible, it is not yet known whether the corresponding theories are sufficiently general to capture the corresponding effective field theory. These cautionary remarks aside, we shall proceed to study the limiting curvature hypothesis in QT gravity minimally coupled to matter, bearing in mind that there may be important differences from the ideal scenario.\footnote{For example, if one considered a resummation of Einstein-Maxwell effective field theory, it would be natural to expect boundedness of certain physical properties of the matter sector as well, \emph{e.g.,} the field strength---see~\cite{Cano:2020qhy, Cano:2020ezi} for an example. }


We will focus on static stars made of an incompressible perfect fluid, which were already explored in Sec.~\ref{Sec_QT_Constant_density}. In this case, the most compact stars are  given by infinite-central-pressure configurations---for those, the curvature at the center diverges. Thus, it is clear the compactness bound given by the limiting curvature hypothesis should be smaller than the divergent-central-pressure one. 



We can inquire about the region in the space of stars such that the Kretschmann scalar in the interior remains smaller than its maximum value in vacuum, $K^{\rm vac}_{\rm max}$.
For constant-density stars in an generic QT theory, the Kretschmann scalar reads
\begin{equation}
\begin{aligned}
    K^{\rm star}&=2D(D-1)\psi_0^2 \\ & \quad\times\frac{\Delta^2-2c_1\Delta\sqrt{\frac{f(R)}{f(r)}}+c_2\frac{ f(R)}{f(r)}}{\left(\Delta-\frac{(D-1)}{(D-3)}\sqrt{\frac{f(R)}{f(r)}}\right)^2} \,, 
\end{aligned}
\end{equation}
where $c_1\equiv \frac{(D-2)}{D}\frac{(D-1)}{(D-3)}$, $ c_2\equiv c_1\frac{(D-1)}{(D-3)}$, and the remaining quantities follow the notation of Sec.~\ref{Sec_QT_Constant_density}. This makes explicit that $ K^{\rm star}$ diverges for stars within the purple region of the previous plots.
Since it is a decreasing function of $r$, the maximum curvature in the interior of a constant-density star occurs at the star's center. Therefore, the boundary in the space of stars between stars which satisfy the vacuum curvature limit and those which do not, is determined by the star radius $R$ for which
\begin{equation} \label{eq:markovlimit}
    K^{\rm star}(r=0)=K^{\rm vac}_{\rm max} \,.
\end{equation}
Although this limit does not admit an analytical expression, it has been depicted as a blue line in previous examples of constant-density plots for Hayward-like stars in Figs.~\ref{fig_D4} and~\ref{fig_D56}. The blue regions correspond to stars which violate the vacuum bounds. Note also that the ``Markov limits'' intersect with multiple constant-central-pressure curves. Therefore, from this point of view, there is no specific value of $\pc$ which plays the role of a Markov bound.


A possible way of avoiding static stars with  curvatures greater than the corresponding vacuum Markov limits consists of imposing the dominant energy condition on the stress tensor. As we show in the plots above, the set of stars compatible with this condition is smaller and seems to be fully contained within the set of stars which do satisfy the vacuum Markov limits. Only for stars very close to the ``critical point,'' depicted in orange in the plots, it is not completely clear whether or not this property is fulfilled. If it is, then there actually exists a value of the central pressure which plays the role of a universal maximum, namely, $\pc = \vrho_0$, which corresponds to the DEC line in the plots.

\section{Conclusions}
\label{sec:disc}



The Buchdahl bound is a powerful constraint on compact objects in GR. With minimal assumptions, one can arrive at a universal bound on the compactness of spherical stars. Remarkably, such stars can be \textit{extremely} compact, with the Buchdahl bound residing comfortably within a photon sphere,\footnote{In Appendix~\ref{app:photonring} we argue that this is also the case for constant-density stars in QT theories.} but not \textit{arbitrarily} compact as there exists a ``compactness gap'' between the most compact star and the Schwarzschild radius. Here we have extended Buchdahl’s analysis to theories that, in vacuum, possess regular geometries (black holes and solitons) as the unique solutions.

We have shown that Buchdahl’s compactness bounds admit a rich and nontrivial generalization in higher-curvature theories whose vacuum sector is free of singularities. For QT gravities admitting regular black holes, we derived the stellar structure equations for generic perfect fluids and obtained analytic control over constant-density configurations. In this setting, the space of solutions is bounded by three distinct limiting regimes: stars with divergent central pressure, configurations with vanishing pressure, and stars whose radius coincides with the inner horizon of the corresponding regular black hole. 

Interestingly, all these bounds intersect at the mass and radius values of the extremal black hole. In the vicinity of this \emph{critical point}, all branches of solutions converge, and although the stellar radius and density can be very similar, the pressure can be radically different, both in magnitude and in sign. Along these lines, we highlight that when higher-curvature terms dominate the dynamics, the static stars must have negative pressures to balance gravity. In fact, in Figs.~\ref{fig_D4} and~\ref{fig_D56} we observe that exotic stars (with negative pressure) appear either for large values of the higher-curvature couplings $\alpha$ compared to the squared Schwarzschild radius or are covered by a horizon, thus corresponding to black holes from the perspective of exterior observers.

For more general matter profiles with monotonically decreasing mean density within a specific range of densities, we established a generalized Buchdahl inequality showing that maximal compactness, for each density profile, is again attained when the interior geometry becomes singular. Interestingly, constant-density stars do not have a universal value for compactness, as it increases for higher densities. Nevertheless, after imposing further assumptions for the theories under consideration---including an upper bound for the maximum value allowed for the mean density---, it can be proven that the subsequent absolute Buchdahl bound for compactness is indeed attained for a specific constant-density star.

Also, despite the existence of universal curvature bounds in vacuum, we found that ordinary matter stars can generically violate these bounds, reaching arbitrarily large curvatures unless additional restrictions---such as the dominant energy condition---are imposed. These results demonstrate that regularity of the vacuum sector alone is insufficient to control the interior structure of ultra-compact stars, and that matter couplings play a decisive role in extending limiting curvature principles beyond black hole solutions. In a follow-up paper~\cite{QTvaccine}, we derive more general conditions under which boundedness of curvature invariants hold even in the presence of matter. As it turns out, the examples discussed here (including constant density stars) represent a somewhat special case, and QT gravity has the ability to yield nonsingular geometries even when the matter sector is severely singular.

The main limitation of our study has been the restriction to minimally coupled matter. The gravitational sector of the theories we have considered includes an infinite tower of higher-order terms. In fact, in the cases with $D \ge 5$, those terms are sufficiently general to describe vacuum effective field theory for gravity. By the same chain of reasoning, the most natural thing to do would be to begin with a matter model ({\it e.g.,} scalar, vector, etc.) and construct theories that include all-orders corrections in the gravity and matter sectors along with nonminimal couplings between them. Constructing theories of this nature with any level of generality is a challenging problem to begin with, and the possibility of studying compact objects like stars in such theories is even more obscure. Hence, our restriction to minimally coupled matter has been for pragmatic reasons. Nonetheless, it is an important problem for future work to understand to what extent our results here capture the physics of more complete models. Along these lines, one could begin by exploring electromagnetic Quasi-topological gravities~\cite{Cano:2020qhy} or theories constructed based on lower-dimensional limits of the Lovelock invariants~\cite{Fernandes:2025fnz, Fernandes:2025mic}.

Even within the setting of QT gravity minimally coupled to matter, there remain interesting problems to explore. For example, the assumptions we have made here are in line with those originally made by Buchdahl. However, modern perspectives on this work often relax or alter several of the underlying assumptions. For example, generalizations of the Buchdahl bound to cases with anisotropic pressures and energy conditions are analytically tractable in GR~\cite{Andreasson:2007ck}, and it would be interesting to extend such considerations to our context. One could also consider Buchdahl bound for charged stars~\cite{Andreasson:2008xw}, or for stars in asymptotically (anti) de Sitter backgrounds~\cite{Boehmer:2005sm} in QT gravity. More generally, there exist other known pressure/density profiles for which the equations are integrable and for which generalized bounds may be obtained, {\it e.g.},~\cite{Buchdahl:1981}.



As was demonstrated in~\cite{Bueno:2024zsx, Bueno:2025zaj}, the spherically symmetric sectors of QT gravities reduce to particular instances of two-dimensional Horndeski theories. From this perspective, our results can be interpreted as bounds on the compactness of two-dimensional stars in these scalar-tensor theories. However, the class of two-dimensional Horndeski theories is much broader than the set of theories singled out by QT gravities~\cite{Kunstatter:2015vxa, Carballo-Rubio:2025ntd}. A natural future direction would be an exploration of Buchdahl-like bounds within the more general class of two-dimensional Horndeski models. It would also be interesting to establish necessary and/or sufficient conditions for the validity of Markov's limiting curvature hypothesis in that setting, both in vacuum and with coupling to matter. These results may extend to higher-dimensional purely gravitational theories as well, provided higher-dimensional theories with these particular Horndeski Lagrangians as their spherical reduction could be identified. Along these lines we note that any such theory outside the Quasi-topological class would necessarily be of a non-polynomial nature.

\section*{Acknowledgements}
We wish to thank Pablo A. Cano for useful discussions. AVC thanks Pau Sol\'e Vilar\'o and Pedro Taranc\'on-\'Alvarez for helpful comments. PB was supported by a Ram\'on y Cajal fellowship (RYC2020-028756-I), by a Proyecto de Consolidación Investigadora (CNS 2023-143822) from Spain’s Ministry of Science, Innovation and Universities, and by the grant PID2022-136224NB-C22, funded by MCIN/AEI/ 10.13039/501100011033/FEDER, UE. \'AJM  was supported by a Juan de la Cierva contract (JDC2023-050770-I) from Spain’s Ministry of Science, Innovation and Universities. AVC was supported by a scholarship of the CEX-2019-000918 project for ``Unidades de excelencia Mar\'ia de Maeztu'' funded by the Spanish Research Agency, AEI/10.13039/501100011033. The authors also acknowledge financial support from the grant CEX2024-001451-M funded by MICIU/AEI/10.13039/501100011033. 

\appendix

\section{Photon sphere radius} \label{app:photonring}
A \emph{photon sphere}  is a spherical surface around a compact object where photons follow unstable circular orbits. The existence of such surfaces is a quintessential feature of strong gravitational fields. For a static, spherically symmetric metric, the radius $r_{\rm \gamma}$ of a photon sphere satisfies
\begin{equation} \label{eq_photon}
    f(r_{\rm \gamma}) = \frac{1}{2} r_{\rm \gamma} \left.\frac{\dif f(r)}{\dif r}\right\vert_{r=r_{\rm \gamma}} \,.
\end{equation}
We can write the derivative of $f(r)$ in terms of $\psi$, $h(\psi)$ and its derivative as follows
\begin{equation}
    \frac{\dif f(r)}{\dif r} = \frac{r}{h'(\psi)} \left[(D-1)h-2\psi h'\right] \,.
\end{equation}
Combining these expressions and substituting $r^2=(1-f)/\psi$, one finds that the radius of the photon sphere must satisfy
\begin{equation}
    f(r_{\rm \gamma}) = \left.\frac{1}{D-1}\frac{(D-1)h-2\psi h'}{h} \right\vert_{r=r_{\rm \gamma}}\,.
\end{equation}
This expression does not allow for an analytic solution for an arbitrary theory. Now, the most compact photon sphere for a given star will occur when its radius coincides with that of the photon sphere, $r_{\rm \gamma}$. For such a star, the RHS of the previous expression depends only on the total density of the star. Moreover, by identifying the effective mean density in the numerator, it can be written as
\begin{equation}
    f(r_{\rm \gamma}) = \left.\frac{D-3}{D-1}\frac{\bvrho\eff}{\bvrho} \right\vert_{r=r_{\rm \gamma}} \,.
\end{equation}
Since $r_\gamma$ is the star radius, this simplifies to
\begin{equation}
    f(r_{\rm \gamma}) = \frac{D-3}{D-1} \left.\Delta\right\vert_{r=r_{\rm \gamma}} \,,
\end{equation}
which is exactly the square root of $f(R_{\rm min})$ for constant density stars---see~\eqref{eq_buchdahlcondition}. Let us recall that $f(r)$ is monotonically increasing when~\eqref{eq_creciente} holds---which also corresponds to positive $\Delta$---, and describes the exterior in the black hole solutions and the ``farthest region'' in the soliton solutions. In both cases, $0<f(r)<1\,$. Because $\Delta$ is an injective function of the density, the solution is unique for a given constant density. However, this uniqueness does not hold when expressed in terms of the mass. Similarly to the divergent-central-pressure behavior, the photon radius is unique for black hole solutions, and for masses smaller than the critical one, there exist two photon radii. In the former cases, since $f(r_{\rm \gamma})>f(R_{\rm min})$, we conclude that, for constant-density stars, the divergent-central-pressure limit lies inside the photon sphere. Consequently, any other more compact star (if existing) would also lie within the photon sphere. Therefore, it follows that the Buchdahl limit for compactness of stars in QTs is always contained in the photon sphere.



\bibliographystyle{JHEP-2}
\bibliography{Gravities}
\noindent 

\end{document}